\documentclass{jfm}
\usepackage{amsmath}
\usepackage{graphicx}
\usepackage{epstopdf, epsfig}
\usepackage{amsfonts}
\usepackage{amssymb}
\usepackage{xcolor}
\usepackage{bm}
\usepackage[version=4]{mhchem}

\newcommand{\PP}{\mathcal{P}}

\shorttitle{Lattice Boltzmann model for reactive mixtures}
\shortauthor{N. Sawant, B. Dorschner and I. V. Karlin}

\title{Consistent lattice Boltzmann model for reactive mixtures}

\author{N. Sawant\aff{1},
  B. Dorschner\aff{1}
 \and I. V. Karlin\aff{1}\corresp{\email{ikarlin@ethz.ch}}}

\affiliation{\aff{1}Department of Mechanical and Process Engineering, ETH Zurich, 8092 Zurich, Switzerland}

\begin{document}

\maketitle

\begin{abstract}
A new lattice Boltzmann model (LBM) for chemically reactive mixtures is presented. The approach capitalizes on the recently introduced thermodynamically consistent LBM for multicomponent mixtures of ideal gases. Similar to the non-reactive case, the present LBM features Stefan--Maxwell diffusion of chemical species and a fully on-lattice mean-field realization of the momentum and energy of the flow. Besides introducing the reaction mechanism into the kinetic equations for the species, the proposed LBM also features a new realization of the compressible flow by using a concept of extended equilibrium on a standard lattice in three dimensions. The full thermodynamic consistency of the original non-reactive multicomponent LBM enables to extend the temperature dynamics to the reactive mixtures by merely including the enthalpy of formation in addition to the previously considered sensible energy. 
Furthermore, we describe in detail the boundary conditions to be used for reactive flows of practical interest. 
The model is validated against a direct numerical simulation of various burning regimes of a hydrogen/air mixture in a microchannel, in two and three dimensions. Excellent comparison in these demanding benchmarks indicates that the proposed LBM can be a valuable and universal model for complex reactive flows.
\end{abstract}


\tableofcontents

\section{Introduction}
\label{sec:intro}

The lattice Boltzmann method (LBM) models fluid flow using a fully discrete kinetic system of designer particles with discrete velocities $\bm{c}_i$, 
fitting into a regular space-filling lattice. In the LBM, the kinetic evolution equation for the populations $f_i(\bm{x},t)$ follows a simple algorithm of ``stream along links $\bm{c}_i$ and collide at the nodes $\bm{x}$ in discrete time $t$". 
Since its inception \citep{higuera_boltzmann_1989,higuera_simulating_1989}, LBM has evolved into a versatile tool for the simulation of complex flows including but not limited to turbulent flows
\citep{dorschner_entropic_2016,dorschner_transitional_2017}, compressible flows  \citep{frapolli_entropic_2016,dorschner_particles_2018,xu_analysis_2013,yang_coupled_2018,lin_mrt_2018}, multiphase flows \citep{mazloomi_entropic_2015,mazloomi_drops_2017,wohrwag_ternary_2018}, rarefied gas \citep{shan_kinetic_2006} and nanoflow \citep{montessori_effects_2016,montemore_effect_2017}.
While the majority of the LBM development concerns single-component fluids, the case of mixtures, and especially of reactive mixtures, remains an active area of research \citep{yan_lattice_2013,lin_mrt_2018,hosseini_mass-conserving_2018,hosseini_hybrid_2019,hosseini_low-mach_2020, feng_lattice-boltzmann_2018,tayyab_experimental_2020,tayyab_lattice-boltzmann_2021}.

Recently, in \citet{sawant_consistent_2021}, we revisited the LBM construction for a compressible multicomponent mixture, focusing on a thermodynamically consistent coupling between diffusion and momentum and energy transfer. 
The species kinetic equations recovered the Stefan--Maxwell diffusion with barodiffusion in the hydrodynamic limit. 
In addition, we also validated and derived from our kinetic model approximate diffusion models such as Curtiss--Hirschfelder and generalized Fick \citep{kee_chemically_2003,poinsot_theoretical_2005,giovangigli_multicomponent_2012}.
A mean-field was introduced for the lattice Boltzmann formulation of the mixture momentum and energy using a  two-population lattice Boltzmann equation for the mixture. The mean-field approach consists of two lattice Boltzmann equations, one for the  mixture density and momentum and another one for the energy with the help of a modification of the non-equilibrium fluxes. 
The two-population mixture LBM and the lattice Boltzmann scheme for the species kinetic equations were realized on the standard three-dimensional lattice. The resulting LBM provides a reduced description of the $M$-component mixture with $M+2$ tightly coupled lattice Boltzmann equations.

In this paper, we extend the two-population mixture LBM to reactive flows and show viability and accuracy for practical applications. 
To that end, we propose novel boundary conditions for walls as well as inlets and outlets. 
Furthermore, unlike previous realizations \citet{sawant_consistent_2021,sawant_lattice_2021}, 
we use the extended lattice Boltzmann method \citep{saadat_extended_2021} for the mean field model.
For validation, we start with simulations of a perfectly stirred reactor and a one dimensional laminar flame.  
Subsequently, combustion of a lean hydrogen/air mixture in microtubes is simulated in two and three dimensions. 
A variety of different flame dynamics such as the periodic ignition-extinction, stable symmetric V-shaped flames and asymmetric flames are captured by the reactive LBM and our simulations are in quantitative agreement with the direct numerical simulations (DNS)  of \citet{pizza_dynamics_2008-1,pizza_three-dimensional_2010}. 
This demonstrates that the proposed is a viable alternative for the simulation of reactive flows.

The paper is structured as follows: We begin with a recap of the nomenclature and the kinetic system for the species in section \ref{sec:stefanMaxwell}. 
This section presents the discrete lattice Boltzmann equations for the reactive species and their implementation on the standard lattice. 
The section closes with a short discussion on time integration of the reaction mass source term. 
Next, we turn our attention to describing the mean field approach for modeling the momentum and energy of the reactive mixture in section \ref{sec:lbMixtures}. 
Here, we discuss the adoption of the extended lattice Boltzmann method for reactive flows and present the realization on standard lattice with the two-population approach. 
The section closes with a brief outline of the resultant macroscopic Navier-Stokes equations in the continuum limit followed by validation with the perfectly stirred reactor and one dimensional laminar flame. 
Having completely described the dynamics in bulk of the multicomponent fluid, we proceed to formulate the boundary conditions for the combined model in section \ref{sec:bc}. 
In section \ref{sec:wallbc}, we discuss the equivalent of no-slip adiabatic boundary condition using the popular bounce-back method. 
This is followed by a technique to implement an isothermal wall based on the  Tamm-Mott-Smith boundary condition in the lattice Boltzmann framework. 
Next, a realization for applying the inlet flux boundary condition is discussed in section \ref{sec:inletbc}. 
In the section \ref{sec:outlet}, a convective boundary condition, which can be used in conjunction with the characteristics based boundary condition, is provided to approximate the species mass fractions at the outlet. 
With the resultant model, we compute the combustion of a premixed hydrogen/air mixture flowing through hot microtubes in section \ref{sec:results}. Validation is performed in different regimes by changing the inlet velocity. The 2D simulation exhibits rich flame dynamics by undergoing repetitive extinction-ignition, forming stable V-shaped flame and stable asymmetric flames. Finally, a 3D open flame is computed in a  microtube. 
\section{Lattice Boltzmann model for the species}
\label{sec:stefanMaxwell}
\subsection{Kinetic equations for the species}
The nomenclature follows \cite{sawant_consistent_2021}. 
The composition of a reactive mixture of $M$ components is described by the species densities $\rho_a$, $a=1,\dots, M$, while the mixture density is,
\begin{align}
\rho=\sum_{a=1}^{M}\rho_a.
\label{eq:rhoSpec}
\end{align}
The rate of change of species densities due to reaction, $\dot \rho_a^{\rm r}$, satisfies mass conservation,
 \begin{equation}
 \sum_{a=1}^{M}\dot \rho_a^{\rm r} = 0.
 \label{eq:sumRhoDot}
 \end{equation}
Introducing the mass fraction,
$Y_a={\rho_a}/{\rho}$, the  molar mass of the mixture $m$ is given by
$
{m}^{-1}=\sum_{a=1}^M Y_a/m_a, 
$
where $m_a$ is the molar mass of the component $a$.
The ideal gas equation of state (EoS) provides a relation between the pressure $P$, the temperature $T$ and the composition,
\begin{equation}
P=\rho R T,
\label{eqn:eosIdealGas}
\end{equation}
where $R={R_U}/{m}$ is the specific gas constant of the mixture and $R_U$ is the universal gas constant.
The pressure of an individual component is related to the pressure of the mixture through Dalton's law of partial pressures, 
$P_a=X_a P$, where the mole fraction is $X_a={m} Y_a /{m_a}$.
Combined with the equation of state (\ref{eqn:eosIdealGas}), the partial pressure takes the form $P_a=\rho_a R_a T$,
where $R_a={R_U}/{m_a}$ is the specific gas constant of the component.

In the kinetic representation, each component is described by a set of populations $f_{ai}$ corresponding to the discrete velocities $\bm{c}_i$, $i=0,\dots, Q-1$. The 
species densities $\rho_a$ and the partial momenta $\rho_a \bm{u}_a$ are defined accordingly,
\begin{align}
&	\rho_a= \sum_{i=0}^{Q-1}f_{ai},\label{eq:density1}\\
&	\rho_a \bm{u}_a= \sum_{i=0}^{Q-1} f_{ai}\bm{c}_i,
	\label{eq:momentum1}
\end{align}
while partial momenta sum up to the mixture momentum, 
\begin{align}
	\rho\bm{u}=\sum_{a=1}^M\rho_a\bm{u}_a.
	\label{eq:totmomSpec}
	\end{align}
Following \cite{sawant_consistent_2021,sawant_lattice_2021}, the kinetic equations for the species can be written as,
\begin{equation}
\partial_t f_{ai} + \bm{c}_{i}\cdot \nabla f_{ai} = \sum_{b\ne a}^M \frac{PX_aX_b}{\mathcal{D}_{ab}} \left[ \left( \frac{f_{ai}^{\rm eq}-f_{ai}}{\rho_a} \right) - \left( \frac{f_{bi}^{\rm eq}-f^*_{bi}}{\rho_b} \right) \right]+ \dot f_{ai}^{\rm r},
\label{eqn:stefanMaxwell} 
\end{equation}
where  $\mathcal{D}_{ab}$ are Stefan--Maxwell binary diffusion coefficients, while the reaction source term satisfies the following conditions, consistent with (\ref{eq:density1}):
\begin{align}
& \sum_{i=0}^{Q-1} \dot f_{ai}^{\rm r}= \dot \rho_a^{\rm r}, \label{eq:Dotfr}\\
&\sum_{i=0}^{Q-1} \dot f_{ai}^{\rm r}\bm{c}_i =  \dot \rho_a^{\rm r} \bm{u}.
 \label{eq:m1fDotc}
\end{align}
We now proceed with specifying the  equilibrium $f_{ai}^{\rm eq}$, the quasi-equilibrium $f^*_{ai}$ and the reaction source term $\dot f_{ai}^{\rm r}$. 
\subsection{Standard lattice and product-form}
Kinetic model (\ref{eqn:stefanMaxwell}) is realized on the standard discrete velocity set $D3Q27$, where $D=3$ stands for three dimensions and $Q=27$ is the number of discrete velocities,
\begin{equation}\label{eq:d3q27vel}
	\bm{c}_i=(c_{ix},c_{iy},c_{iz}),\ c_{i\alpha}\in\{-1,0,1\}.
\end{equation}
In order to specify the equilibrium $f_{ai}^{\rm eq}$, the quasi-equilibrium $f^*_{ai}$ and the reaction source term $\dot f_{ai}^{\rm r}$ in (\ref{eqn:stefanMaxwell}), we first define a triplet of functions in two variables, $\xi_{\alpha}$ and $\PP_{\alpha\alpha}$, 
\begin{align}
&	\Psi_{0}(\xi_{\alpha},\PP_{\alpha\alpha}) = 1 - \PP_{\alpha\alpha}, 
	\label{eqn:phi0}
	\\
&	\Psi_{1}(\xi_{\alpha},\PP_{\alpha\alpha}) = \frac{\xi_{\alpha} + \PP_{\alpha\alpha}}{2},
	\label{eqn:phiPlus}
	\\
&	\Psi_{-1}(\xi_{\alpha},\PP_{\alpha\alpha}) = \frac{-\xi_{\alpha} + \PP_{\alpha\alpha}}{2},
	\label{eqn:phis}
\end{align}
%
and consider a product-form associated with the discrete velocities $\bm{c}_i$ (\ref{eq:d3q27vel}),
\begin{equation}\label{eq:prod}
	\Psi_i= \Psi_{c_{ix}}(\xi_x,\PP_{xx}) \Psi_{c_{iy}}(\xi_y,\PP_{yy}) \Psi_{c_{iz}}(\xi_z,\PP_{zz}).
\end{equation}
All pertinent populations to be encountered in this paper shall be determined by specifying the parameters 
$\xi_\alpha$ and $\PP_{\alpha\alpha}$ in the product-form (\ref{eq:prod}). 
To that end, the equilibrium  and the quasi-equilibrium populations are found by setting,
\begin{align}
	&\xi_\alpha=u_\alpha,\\ 
	&\PP_{\alpha \alpha}=R_aT+u_{\alpha}^2,
	\label{eq:xiPeq}
\end{align}
in the former, and 
\begin{align}
	&\xi_\alpha=u_{a\alpha},\\ 
	&\PP_{\alpha \alpha}=R_aT+u_{a\alpha}^2,
	\label{eq:xiPstar}
\end{align}
in the latter cases:
\begin{align}
	f_{ai}^{\rm eq}
	&= \rho_a\Psi_{c_{ix}}\left(u_x,u_x^2+R_aT\right) \Psi_{c_{iy}}\left(u_y,u_y^2+R_aT\right) \Psi_{c_{iz}}\left(u_z,u_z^2+R_aT\right),
\label{eq:27eq}	\\
	f_{ai}^{*}
	&= \rho_a\Psi_{c_{ix}}\left(u_{ax},u_{ax}^2+R_aT\right)
	\Psi_{c_{iy}}\left(u_{ay},u_{ay}^2+R_aT\right) 
	\Psi_{c_{iz}}\left(u_{az},u_{az}^2+R_aT\right).
\label{eq:27qeq}
\end{align}
Reaction terms  are  specified with the product-form (\ref{eq:prod}) using the equilibrium parameters (\ref{eq:xiPeq}),
\begin{align}
    	\dot f_{ai}^{\rm r}
    	&= \dot \rho_a^{\rm r}  \Psi_{c_{ix}}\left(u_{x},u_{x}^2+R_aT\right)
	\Psi_{c_{iy}}\left(u_{y},u_{y}^2+R_aT\right) 
	\Psi_{c_{iz}}\left(u_{z},u_{z}^2+R_aT\right).
\label{eq:fsource}
\end{align}
Analysis of the hydrodynamic limit of the kinetic model (\ref{eqn:stefanMaxwell}) follows the lines already presented in \cite{sawant_consistent_2021}. 
The balance equations for the densities of the species in the presence of the source term are found as follows,
\begin{align}
	\partial_t\rho_a&=-\nabla\cdot(\rho_a \bm{u})-\nabla\cdot(\rho_a \delta\bm{u}_a) + \dot \rho_a^{\rm r},
	\label{eq:dtrhoa}
\end{align}
where the diffusion velocities, $\delta\bm{u}_a=\bm{u}_a-\bm{u}$, satisfy the Stefan--Maxwell constitutive relation,
\begin{equation}
	P\nabla X_a+(X_a-Y_a)\nabla P=\sum_{b\ne a}^M \frac{PX_aX_b}{\mathcal{D}_{ab}} \left({\delta}\bm{u}_{b} - {\delta}\bm{u}_a \right).
	\label{eq:constit2}
\end{equation}
Summarizing, kinetic model (\ref{eqn:stefanMaxwell}) recovers both the Stefan--Maxwell law of diffusion
and the composition change due to chemical reaction, as presented in equation (\ref{eq:dtrhoa}).

\subsection{Lattice Boltzmann equation for the species}

Derivation of the lattice Boltzmann equation from the kinetic model (\ref{eqn:stefanMaxwell}) proceeds along the lines of the non-reactive case already presented in detail by \cite{sawant_consistent_2021}. 
Upon integration of \eqref{eqn:stefanMaxwell} along the characteristics and application of the trapezoidal rule to all relaxation terms on the right hand side except for the reaction term, we arrive
at a fully discrete lattice Boltzmann equation for the species,
\begin{equation}
	f_{ai}(\bm{x}+\bm{c}_i \delta t, t+ \delta t)  = f_{ai}(\bm{x},t)+ 2 \beta_a [f_{ai}^{\rm eq}(\bm{x},t) - f_{ai}(\bm{x},t)]
	+ \delta t (\beta_a-1) F_{ai}(\bm{x}, t) +  R_{ai}^{\rm r}.
	\label{eqn:finalNumericalEquationsReactive}
\end{equation}
Here $\delta t$ is the lattice time step, the equilibrium populations are provided by Eq.\ (\ref{eq:27eq}), while the relaxation parameters $\beta_a\in[0,1]$ are,
\begin{equation}
	\beta_a=\frac{\delta t}{2 \tau_a + \delta t}.
	\label{eqn:betaa}
\end{equation}
Their relation to the Stefan--Maxwell binary diffusion coefficients is found as follows:
Introducing characteristic times,
\begin{equation}
	\tau_{ab}=\frac{mR_UT}{\mathcal{D}_{ab}m_a m_b},
\end{equation} 
the relaxation times $\tau_a$ in (\ref{eqn:betaa}) are defined through mixture-averaging, 
\begin{equation}
	\frac{1}{\tau_a} = \sum_{b\ne a}^M \frac{Y_b}{\tau_{ab}}.
	\label{eqn:taua}
\end{equation}
Furthermore in (\ref{eqn:finalNumericalEquationsReactive}), the quasi-equilibrium relaxation term $F_{ai}$ is spelled out as follows,
\begin{equation}
F_{ai} = Y_a \sum_{b\ne a}^M \frac{1}{\tau_{ab}}  \left( f_{bi}^{\rm eq}-f_{bi}^* \right).
\label{eqn:fStar}
\end{equation}
Here the quasi-equilibrium populations $f_{bi}^*$ are defined by the product-form (\ref{eq:27qeq}), subject to the following parameterization,
\begin{align}
	&\xi_\alpha=u_{\alpha}+V_{b\alpha},\\ &\PP_{\alpha \alpha}=R_bT+\left(u_{\alpha}+V_{b\alpha}\right)^2,
	\label{eq:xiPstar2}
\end{align}
where the second-order accurate diffusion velocity $\bm{V}_b$ is the result of the lattice Boltzmann discretization of the kinetic equation and is found by solving the $M\times M$ linear algebraic system for each spatial component,
\begin{align}
\left( 1+ \frac{\delta t}{2 \tau_a}\right)  {\bm{V}_{a}} - \frac{\delta t}{2} \sum_{b\ne a}^{M} \frac{1}{\tau_{ab}} Y_b  {\bm{V}_{b}}=\bm{u}_{a}-\bm{u}.
\label{eqn:transform1}
\end{align}
The system (\ref{eqn:transform1}) has been derived in \citet{sawant_consistent_2021} and is not altered by the presence of the reaction. 
In our realization, we solve (\ref{eqn:transform1}) with the Householder QR decomposition method from the Eigen library \citep{guennebaud_eigen_2010}. 

All the elements of the lattice Boltzmann equation (\ref{eqn:finalNumericalEquationsReactive}) described so far are identical to those already present in the non-reactive case of \cite{sawant_consistent_2021}. 
 Finally, the reaction term in (\ref{eqn:finalNumericalEquationsReactive}) is represented by an integral over the characteristics,
\begin{align}
	{R_{ai}^{\rm r}}=\delta t\int_{0}^{1}\dot f_{ai}^{\rm r}(\bm{x}+\bm{c}_i s\delta t,t+s\delta t )ds.
	\label{eq:R1}
\end{align}
Taking into account the structure of the reaction term (\ref{eq:fsource}), we use a simple explicit approximation for the implicit term (\ref{eq:R1}),
\begin{align}
	{R_{ai}^{\rm r}}\approx\rho^{-1}f_{ai}^{\rm eq}(\bm{x},t)\delta t\int_{0}^{1}\dot \rho_{a}^{\rm r}(\bm{x},t+s\delta t)ds.
	\label{eq:R2}
\end{align}
Reaction rates $\dot \rho_a^{\rm r}$ are obtained from the open source chemical kinetics package Cantera \citep{goodwin_cantera_2018} as a function of mixture internal energy $U$ and composition, $\dot \rho_a^{\rm r}=\dot \rho_a^{\rm r}(U,\rho_1,\dots,\rho_M)$. In order to mitigate the stiffness of the reaction rates for  detailed reaction mechanisms, we introduce a time step $\delta t^{\rm r} = {\delta t}/{l}$, where $l=1,2,...$ and evaluate (\ref{eq:R2}) by forward Euler in $l$ sub-steps,
\begin{align}
	R_{ai}^{\rm r}\approx \rho^{-1}f_{ai}^{\rm eq}(\bm{x},t)\left[
	\delta t^{\rm r}\sum_{s=0}^{l-1}\dot \rho_a^{\rm r}\left( U(\bm{x},t),\rho_1\left(\bm{x},t+s\delta t^{\rm r}\right),\dots,\rho_M\left(\bm{x},t+s\delta t^{\rm r}\right)\right)\right].\label{eq:R3}
\end{align}
{Note that, during sub-iterations, the energy remains fixed although the temperature changes, in general. In other words, at each grid point, sub-iterations implement a zero-dimensional perfectly stirred reactor.}
Execution time for sub-steps increases by about $ 6 \%$ for $l=2$ and by $15 \%$ for $l=4$. 
In this paper, we use $l=2$, which is small enough that the integration error does not influence the flow solution but still reduces the computational complexity by roughly half due to the larger time step of the fluid solver, $\delta t = 2 \delta t^r$.

Summarizing, the lattice Boltzmann system (\ref{eqn:finalNumericalEquationsReactive}) delivers the extension of the species dynamics subject to the Stefan--Maxwell diffusion to the reactive mixtures. We now proceed with setting up the lattice Boltzmann equations for the mixture momentum and energy.
\section{Lattice Boltzmann model of mixture momentum and energy}
\label{sec:lbMixtures}
%

\subsection{Double-population lattice Boltzmann equation}
\label{sec:energy}
The mass-based specific internal energy ${U}_{a}$ and enthalpy ${H}_{a}$ of the species are,
\begin{align}
	{U}_{a}&=U^0_a+\int_{T_0}^T{C}_{a,v}(T')dT',
	\label{eq:specUa}\\
	{H}_{a}&=H^0_a+\int_{T_0}^T{C}_{a,p}(T')dT',
	\label{eq:specHa}
\end{align}
where $U^0_a$ and  $H^0_a$ are the energy and the enthalpy of formation at the reference temperature $T_0$, respectively, 
while $C_{a,v}$ and $C_{a,p}$ are specific heats at constant volume and at constant pressure, satisfying the Mayer relation, ${C}_{a,p}-{C}_{a,v}=R_{a}$.
Consequently, the internal energy $\rho U$ and enthalpy $\rho H$ of the mixture are defined as,
\begin{align}
	\rho U=\sum_{a=1}^M\rho_a U_a,
		\label{eq:U}\\
	 \rho H=\sum_{a=1}^M\rho_a H_a.
	\label{eq:H}
\end{align}
While the sensible heat was considered in the non-reactive case \citep{sawant_consistent_2021}, by taking into account the heat of formation we immediately extend the model to reactive mixtures. Same as in \citet{sawant_consistent_2021}, we follow a two-population approach. One set of populations ($f$-populations) is used to represent the density and the momentum of the mixture.
Below, we refer to the $f$-populations as the momentum lattice.  The locally conserved fields are the density and the momentum of the mixture,
\begin{align}
&\sum_{i=0}^{Q-1} f_i  = \rho,\label{eqn:fdensity}\\
&\sum_{i=0}^{Q-1} f_i \bm{c}_{i} = \rho \bm{u}.
\label{eqn:f1momMomentum} 
\end{align}
Another set of populations ($g$-populations), or the energy lattice, is used to represent the local conservation of the total energy of the mixture,
\begin{align}
&\sum_{i=0}^{Q-1} g_i  =  \rho E,  \label{eqn:g0momTotalEnergy} \\
&\rho E=\rho U + {\frac{\rho u^2}{2}}.
\label{eq:totalE}
\end{align}
Since the mixture internal energy (\ref{eq:U}) depends on the composition, the species kinetic equations become coupled with the kinetic equations for the mixture to be introduced shortly. Conversely, the temperature is evaluated by solving the  integral equation, cf.\ (\ref{eq:specUa}) and (\ref{eq:U}),
\begin{equation}
\label{eq:temperature}
    \sum_{a=1}^MY_a\left[U_a^0 + \int_{T_0}^T {C}_{a,v}(T')dT'\right]=E-\frac{u^2}{2}.
\end{equation}
The temperature evaluated by solving (\ref{eq:temperature}) is used as the input in the equation of state (\ref{eqn:eosIdealGas}) elsewhere in the species lattice Boltzmann system. This furnishes a two-way coupling input between the species and the mixture kinetic systems.

Similar to \citet{sawant_consistent_2021}, the lattice Boltzmann equations for the momentum and for the energy lattice are patterned from the single-component developments and are realized on the $D3Q27$ discrete velocity set.
While the prototype single-component LBM used in \citet{sawant_consistent_2021} was that of \cite{saadat_lattice_2019}, here we take advantage of a more recent proposal by \cite{saadat_extended_2021}. It is noted that, while both these single-component models are essentially equivalent, the recent formulation is more compact in its formulation and simpler in terms of implementation.
Following the more recent proposal, the mixture lattice Boltzmann equations are written,
\begin{align}
f_i(\bm{x}+\bm{c}_i \delta t,t+\delta t)- f_i(\bm{x},t)&=  \omega (f_i^{\rm ex} -f_i),  \label{eqn:f} 
\\
g_i(\bm{x}+\bm{c}_i \delta t,t+ \delta t) - g_i(\bm{x},t)&=  \omega_1 (g_i^{\rm eq} -g_i) + (\omega - \omega_1) (g_i^{*} -g_i),
 \label{eqn:g}
\end{align}
where relaxation parameters $\omega$ and $\omega_1$ are related to the mixture viscosity and thermal conductivity, and we proceed with specifying the pertinent populations in (\ref{eqn:f}) and (\ref{eqn:g}).

\subsection{Extended equilibrium for the momentum lattice}
\label{sec:ExtendedEquilibriumMmomentum}

The extended equilibrium populations $f_i^{\rm ex}$ in (\ref{eqn:f}) are specified by the product-form (\ref{eq:prod}), 
with the parameters identified  as ${\xi}_{\alpha}={u}_{\alpha}$ and $\PP_{\alpha \alpha}=\PP_{\alpha \alpha}^{\rm ex}$,
\begin{equation}
f_{i}^{\rm ex}=\rho 
\Psi_{c_{ix}}\left(u_{x},\PP_{xx}^{\rm ex}\right)
\Psi_{c_{iy}}\left(u_{y},\PP_{yy}^{\rm ex}\right) 
\Psi_{c_{iz}}\left(u_{z},\PP_{zz}^{\rm ex}\right),
\label{eq:feqmix}
\end{equation}
where the extended parameter $\PP_{\alpha \alpha}^{\rm ex}$ reads,
\begin{align}
	\PP_{\alpha \alpha}^{\rm ex}& = \PP_{\alpha \alpha}^{\rm eq} + \delta t\left( \frac{2-\omega}{2\rho\omega}\right)
	\partial_\alpha \left(\rho u_\alpha (1 - 3 R T) - \rho u_\alpha^3\right),
	\label{eqn:Pex}
\end{align}
while $\PP_{\alpha \alpha}^{\rm eq}$, 
\begin{align}
	\PP_{\alpha \alpha}^{\rm eq}& = RT+u_\alpha^2,
	\label{eqn:Peq}
\end{align}
corresponds to the conventional product-form equilibrium,
\begin{equation}
	f_{i}^{\rm eq}=\rho 
	\Psi_{c_{ix}}\left(u_{x},\PP_{xx}^{\rm eq}\right)
	\Psi_{c_{iy}}\left(u_{y},\PP_{yy}^{\rm eq}\right) 
	\Psi_{c_{iz}}\left(u_{z},\PP_{zz}^{\rm eq}\right),
	\label{eq:feq}
\end{equation}
The effect of extension, featured by the second term in (\ref{eqn:Pex}), is to correct for the incomplete Galilean invariance of the standard $D3Q27$ velocity set (\ref{eq:d3q27vel}). With the original formulation of the mixture momentum lattice in \citet{sawant_consistent_2021}, a similar correction was achieved by augmenting Eq.\ (\ref{eqn:f}) with an additional forcing term which required evaluation of second-order derivatives in space. In the present formulation, the correction of Galilean invariance is achieved by the extended equilibrium which requires evaluation of only a first-order derivative, cf.\ Eq.\ (\ref{eqn:Pex}), a more local operation.
\subsection{Equilibrium and quasi-equilibrium of the energy lattice}
\label{sec:ExtendedEquilibriumEnergy}

Turning our attention to the energy lattice, the corresponding equilibrium and quasi-equilibrium populations in (\ref{eqn:g}) are evaluated along the lines of \cite{saadat_extended_2021}:
Let us introduce linear operators $\mathcal{O}_\alpha$, acting on any smooth function $A(\bm{u},T)$ according to a rule,
\begin{equation}
    \mathcal O_\alpha A= 
    RT \frac{\partial A}{\partial u_\alpha} + u_\alpha A.
    \label{eqn:oalpha}
\end{equation}
The equilibrium populations $g_i^{\rm eq}$ are specified with an operator version of the product-form (\ref{eq:prod}). To that end, we consider parameters ${\xi_\alpha}$ and $\PP_{\alpha\alpha}$ as operator symbols,
\begin{align}
    &\xi_\alpha = \mathcal{O}_\alpha, \label{eq:Oa}\\
    &\PP_{\alpha\alpha} = \mathcal{O}_\alpha^2. \label{eq:OaOa}
\end{align}
With the operators (\ref{eq:Oa}) and (\ref{eq:OaOa}) substituted into the product form (\ref{eq:prod}), the equilibrium populations $g_i^{\rm eq}$ are compactly written using the energy $E$ as the generating function,
\begin{align}
    g_i^{\rm eq} = \rho \Psi_{c_{ix}}(\mathcal{O}_x,\mathcal{O}_x^2) \Psi_{c_{iy}}(\mathcal{O}_y,\mathcal{O}_y^2) \Psi_{c_{iz}}(\mathcal{O}_z,\mathcal{O}_z^2)E. \label{eq:geq_i}
\end{align}
It is straightforward to verify by a direct computation that the equilibrium (\ref{eq:geq_i}) satisfies the necessary conditions to recover the mixture energy equation as in \cite{sawant_consistent_2021}, namely, the equilibrium energy flux $\bm{q}^{\rm eq}$ and the flux thereof $\bm{R}^{\rm eq}$, 
 \begin{align}
 &\bm{q}^{\rm eq}= \sum_{i=0}^{Q-1}  g_i^{\rm eq} \bm{c}_{i} =  \left(H+\frac{u^2}{2}\right)\rho\bm{u},
 \label{eqn:geq1mom} 
 \\
 &\bm{R}^{\rm eq}=\sum_{i=0}^{Q-1} g_i^{\rm eq} \bm{c}_i\otimes\bm{c}_i =
     \left(H+\frac{u^2}{2}\right) \bm{P}^{\rm eq} + P\bm{u}\otimes\bm{u},
 \label{eqn:geq2mom}
 \end{align}
where $H$ is the specific mixture enthalpy (\ref{eq:H}). 
Finally, the quasi-equilibrium populations $g_i^*$ 
differs from the equilibrium $g_i^{\rm eq}$ by the energy flux only  \citep{karlin_consistent_2013,sawant_consistent_2021,saadat_extended_2021},
	\begin{align}
		g_{i}^*= \left\{\begin{aligned}
			& g_{i}^{\rm eq}+\frac{1}{2}\bm{c}_i\cdot\left(\bm{q}^*-\bm{q}^{\rm eq}\right), &\text{ if } c_i^2=1, & \\ 
			&g_i^{\rm eq}, & \text{otherwise}.&\\
		\end{aligned}\right.
	\label{eq:gstar}	
	\end{align}
were $\bm{q}^*$ is a specified quasi-equilibrium energy flux,
\begin{align}
	\bm{q}^{*} &=\sum_{i=0}^{Q-1} g_i^{*} \bm{c}_{i} =  \bm{q} -  \bm{u}\cdot (\bm{P} - \bm{P}^{\rm eq}) +\bm{q}^{\rm diff}+\bm{q}^{\rm corr}+\bm{q}^{\rm ex}.
	\label{eq:gstareq1mom}
\end{align}
All contributions on the right hand side of (\ref{eq:gstareq1mom}), except for the vector $\bm{q}^{\rm ex}$, were already introduced in \cite{sawant_consistent_2021} and do not alter under the present modifications:
The two first terms in (\ref{eq:gstareq1mom}) maintain a variable Prandtl number and include the energy flux $\bm{q}$ and the pressure tensor $\bm{P}$,
\begin{align}
&	\bm{q}=\sum_{i=0}^{Q-1} g_i \bm{c}_{i},\\
&	\bm{P}=\sum_{i=0}^{Q-1} f_i \bm{c}_{i}\otimes \bm{c}_{i}.
\end{align}
The interdiffusion energy flux $\bm{q}^{\rm diff}$,
\begin{align}
\bm{q}^{\rm diff} =\left(\frac{\omega_1}{\omega-\omega_1} \right) \rho\sum_{a=1}^{M}H_aY_a \bm{V}_a,\label{eq:qdiff}
\end{align}
where the diffusion velocities $\bm{V}_a$ are defined by Eq.\ (\ref{eqn:transform1}), contributes the enthalpy transport due to diffusion, cf.\ \citep{sawant_consistent_2021}.
Moreover, the correction flux $\bm{q}^{\rm corr}$ is required in the two-population approach to the mixtures in order to recover the Fourier law of thermal conduction  \citep{sawant_consistent_2021},
\begin{align}
\bm{q}^{\rm corr}=\frac{1}{2}\left(\frac{\omega_1-2}{\omega_1-\omega}\right) {\delta t}P \sum_{a=1}^M  H_{a} \nabla  Y_a.
\label{eq:corrFourier}
\end{align}
Finally, the term $\bm{q}^{\rm ex}$ in the quasi-equilibrium flux  (\ref{eq:gstareq1mom}) is required for consistency with the extended equilibrium (\ref{eq:feqmix}), and is similar to its single-component counterpart \citep{saadat_extended_2021}. Components of the vector $\bm{q}^{\rm ex}$ follow the structure of (\ref{eqn:Pex}), 
\begin{align}
	{q}_\alpha^{\rm ex} &= - \frac{1}{2}\delta t  u_\alpha
	\partial_\alpha \left(\rho u_\alpha \left(1 - 3 R T\right) - \rho u_\alpha^3\right).
	\label{eqn:qex}    
\end{align}
Spatial derivatives in the correction flux (\ref{eq:corrFourier}) and in the isotropy correction (\ref{eqn:Pex}) and (\ref{eqn:qex}) were implemented using  isotropic lattice operators \citep{thampi_isotropic_2013}. 
\subsection{Mixture mass, momentum and energy equations}
With the equilibrium and quasi-equilibrium populations specified, the hydrodynamic limit of the two-population lattice Boltzmann system (\ref{eqn:f}) and (\ref{eqn:g}) is found by expanding the propagation to second order in the time step $\delta t$ and evaluating the moments of the resulting expansion. Analysis is standard, details can be found in  \citet{sawant_consistent_2021} and \citet{saadat_extended_2021}, here we present the final result.
The continuity, the momentum and the energy equations for a reactive multicomponent mixture \citep{williams_combustion_1985,bird_transport_2007} are, respectively,
\begin{align}
&\partial_t \rho + \nabla\cdot (\rho \bm{u})=0,
\label{eqn:dtrho}
\\
&\partial_t (\rho\bm{u}) +  \nabla\cdot ({\rho\bm{u}\otimes\bm{u} })+ \nabla\cdot \bm{\pi}=0,
\label{eqn:dtu}
\\
&\partial_t (\rho E)+\nabla\cdot(\rho E\bm{u})+\nabla\cdot\bm{q}+\nabla\cdot(\bm{\pi}\cdot\bm{u})=0.
\label{eqn:dtE}
\end{align}
Here, the pressure tensor $\bm{\pi}$ in the momentum equation reads,
\begin{equation}\label{eq:NSmix}
\bm{\pi}=P\bm{I}
-\mu \left( \nabla\bm{u}  + \nabla\bm{u}^{\dagger}  -\frac{2}{D} (\nabla\cdot\bm{u})\bm{I} \right) 
-\varsigma (\nabla\cdot\bm{u}) \bm{I},
\end{equation}
where the dynamic viscosity $\mu$ and the bulk viscosity $\varsigma$ are related to the relaxation parameter $\omega$,
\begin{align}
\mu  &= \left( \frac{1}{\omega} - \frac{1}{2}\right) P{\delta t},
\label{eq:mu}\\
\varsigma &=\left( \frac{1}{\omega}-\frac{1}{2}\right)\left( \frac{2}{D} - \frac{R}{C_v} \right) P{\delta t}.
\label{eqn:varsigma}
\end{align}
Here $C_v=\sum_{a=1}^M Y_a C_{a,v}$ is the mixture specific heat at constant volume. The heat flux $\bm{q}$ in the energy equation (\ref{eqn:dtE}) reads,
\begin{equation}
\label{eq:qneq}
\bm{q}=-\lambda\nabla T+\rho\sum_{a=1}^{M}H_aY_a \bm{V}_a .
\end{equation}
The first term in (\ref{eq:qneq}) is the Fourier law of thermal conduction, with thermal conductivity $\lambda$ related to the relaxation parameter $\omega_1$,
\begin{equation}\label{eq:lambda}
\lambda= \left(\frac{1}{\omega_1} - \frac{1}{2}\right) P C_p{\delta t},
\end{equation}
where $C_p=C_v+R$ is the mixture specific heat at constant pressure.
The second term in (\ref{eq:qneq}) is the interdiffusion energy flux. With the thermal diffusivity $\alpha=\lambda/\rho C_p$ and the kinematic viscosity $\nu=\mu/\rho$, the Prandtl number becomes, 
${\rm Pr} = {\nu}/{\alpha}$.
For this reactive formulation, the local dynamic viscosity $\mu(\bm{x},t)$ and the thermal conductivity $\lambda(\bm{x},t)$ of the mixture is evaluated as a function of the local chemical state by using the chemical kinetics solver Cantera \citep{goodwin_cantera_2018}. Cantera employs a combination of methods such as interaction potential energy functions \citep{kee_chemically_2003}, hard sphere approximations, the methods described in \citet{wilke_viscosity_1950} and \citet{mathur_thermal_1967} to calculate the mixture transport coefficients. 

In summary, by virtue of thermodynamic consistency of the  lattice Boltzmann model for mixture momentum and energy \citep{sawant_consistent_2021}, the extension to the reactive case requires merely an upgrade of the sensible heat by the heat of formation. The proposed realization also takes into account the revised formulation of the two-population LBM for compressible flow \citep{saadat_extended_2021}. We proceed to finalizing the model development by specifying the coupling between the lattice Boltzmann models for the species and the mixture momentum and energy, as well as the coupling to the external chemical kinetics solver.

\subsection{Coupling between the species and the mixture subsystems}
With the two subsystems, the species and the mixture, first  constructed  independently from each other and after that being coupled weakly in the way described in \citet{sawant_consistent_2021}, we are left with two independent definitions of the mixture density and the mixture momentum: On the one hand,
the mixture density $\rho$ (\ref{eqn:fdensity}) and the mixture momentum $\rho\bm{u}$ (\ref{eqn:f1momMomentum}) are defined as the moments of the $f$-populations on the momentum lattice.  
On the other hand, the same quantities are defined with the species populations as the sum of partial densities and partial momenta. 
The number of the conservation laws for the species subsystem is $M+D$, while for the mixture subsystem it is $D+2$. 
The total number of the conservation laws in the weakly coupled combined system is $M+2D+2$. Thus, the weakly coupled system is in excess of $D+1$ conservation laws.
%
	This redundancy is eliminated by removing one set of species populations (here, the $M^{th}$) 
	and writing,
	\begin{equation}\label{eq:strongest}
		f_{Mi}=f_{i}-\sum_{a=1}^{M-1}f_{ai}.
	\end{equation}
	As a consequence, the $M^{th}$ component is not an independent field anymore but is slaved to the remaining species and mixture populations. 
The number of independent conservation laws in the resulting strongly coupled system is 
$M+D+1$, which corresponds to the locally conserved fields, $\rho_1,\dots,\rho_{M-1}$ (\ref{eq:density1}),
$\rho$ (\ref{eqn:fdensity}), $\rho\bm{u}$ (\ref{eqn:f1momMomentum}) and $\rho E$ \eqref{eqn:g0momTotalEnergy} . While the assignment of the slaved
component $M$ is not unique, it is advisable to select the component which
carries the majority of mass in the mixture. 
The coupling \eqref{eq:strongest} reduces the number of lattices from $M+2$ to $M+1$.

\subsection{Coupling between lattice Boltzmann and chemical kinetics}
 \label{sec:cantera}
The lattice Boltzmann code is coupled to the open source code chemical kinetics solver Cantera \citep{goodwin_cantera_2018}. The Cantera solver is supplied with the publicly accessible GRI-Mech 3.0 mechanism \citep{smith_gri-mech_1999} as an input. The communication between the lattice Boltzmann solver and the Cantera is summarized as follows: 
\begin{enumerate}
    \item \noindent During the collision step, the lattice Boltzmann solver provides internal energy, specific volume and mass fractions to set the chemical state in Cantera. 
    \item \noindent Cantera numerically solves the integral equation (\ref{eq:temperature}) to find the temperature at that state. 
    \item \noindent The production rates of species $\dot \rho_a^{\rm r}$, transport coefficients including dynamic viscosity, thermal conductivity and the Stefan--Maxwell diffusivities are obtained from Cantera as a function of the current state.
    \item  \noindent In the lattice Boltzmann solver, the temperature is used to evaluate the equilibrium and quasi-equilibrium moments and populations. The transport coefficients are used to calculate the corresponding relaxation times.  
\end{enumerate}
Other thermodynamic parameters necessary for the simulation such as the specific heats and molecular masses are also obtained through Cantera. The reference standard state temperature is $T_0=298.15\, \text{K}$ and the reference standard state pressure is $P_0=1\ \text{atm}$. The data required by the lattice Boltzmann solver during runtime is obtained by querying Cantera through its C++ API. In all cases considered in this paper, we use the detailed  mechanism of hydrogen/air combustion \citep{li_updated_2004} involving  nine species, \ce{N2}, \ce{O2}, \ce{H2}, \ce{H}, \ce{O}, \ce{OH}, \ce{H2O}, \ce{HO2} and \ce{H2O2}. Finally, same as in \citet{sawant_consistent_2021}, acoustic scaling is used for conversion of length and time between the physical and the lattice units. The speed of sound at a specified reference composition and specified temperature (typically, at the unburnt mixture state) is used to make the velocity non-dimensional. The characteristic length in the respective setup is used to rescale the length.

We shall now proceed with a validation of the coupled reactive flow lattice Boltzmann model in two test cases.
The perfectly stirred reaction (PSR) simulation is selected to validate the multistep approach to the evaluation of the reaction term (\ref{eq:R3}) while the laminar flame speed simulation is to probe the coupling of the new formulation of the mixture momentum and energy LBM of sec.\  \ref{sec:ExtendedEquilibriumMmomentum} and \ref{sec:ExtendedEquilibriumEnergy}. 

\subsection{Perfectly stirred reactor}
\label{sec:PSR}

\begin{figure}
	\centering
\includegraphics[width=0.8\linewidth]{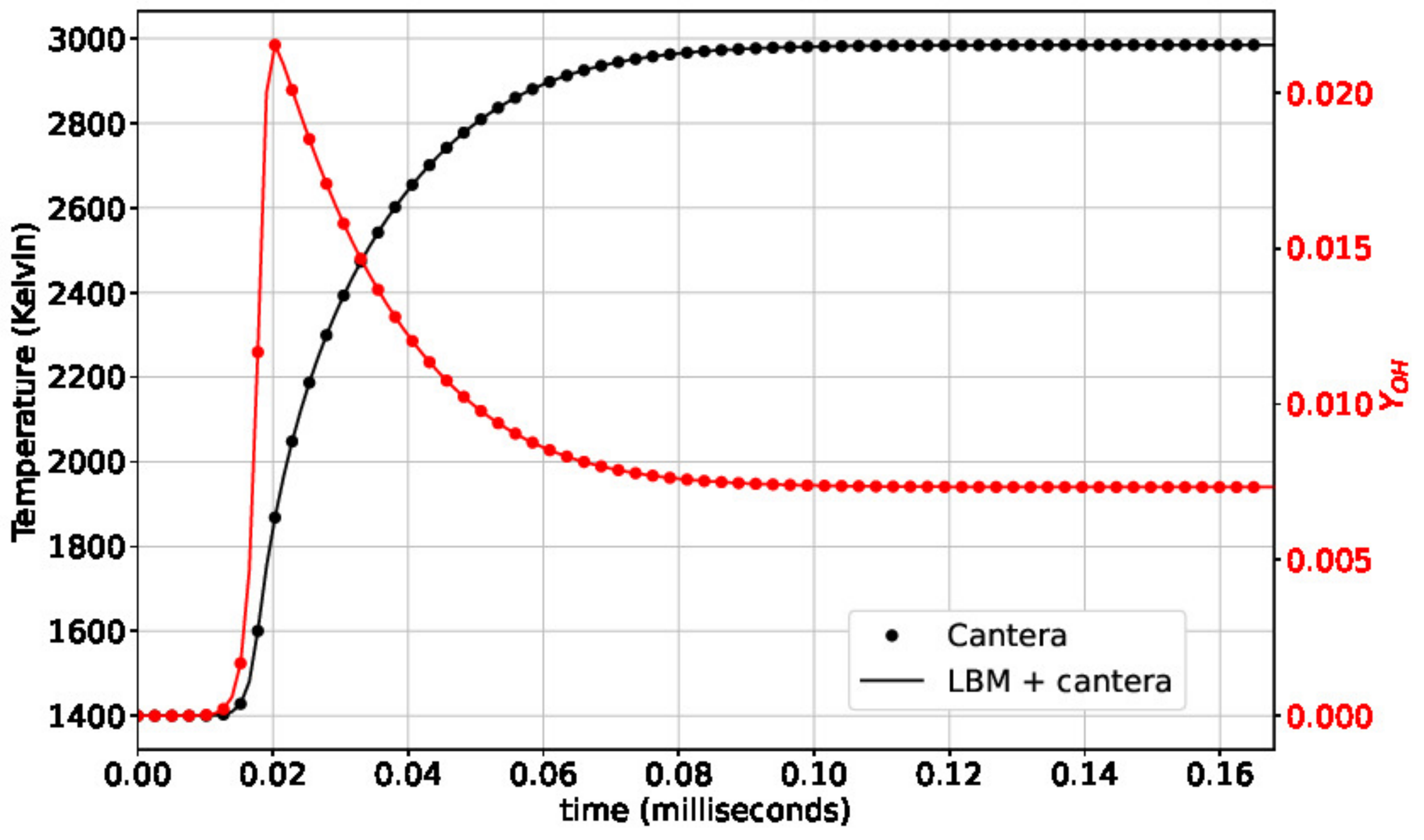} \\
\includegraphics[width=0.7\linewidth]{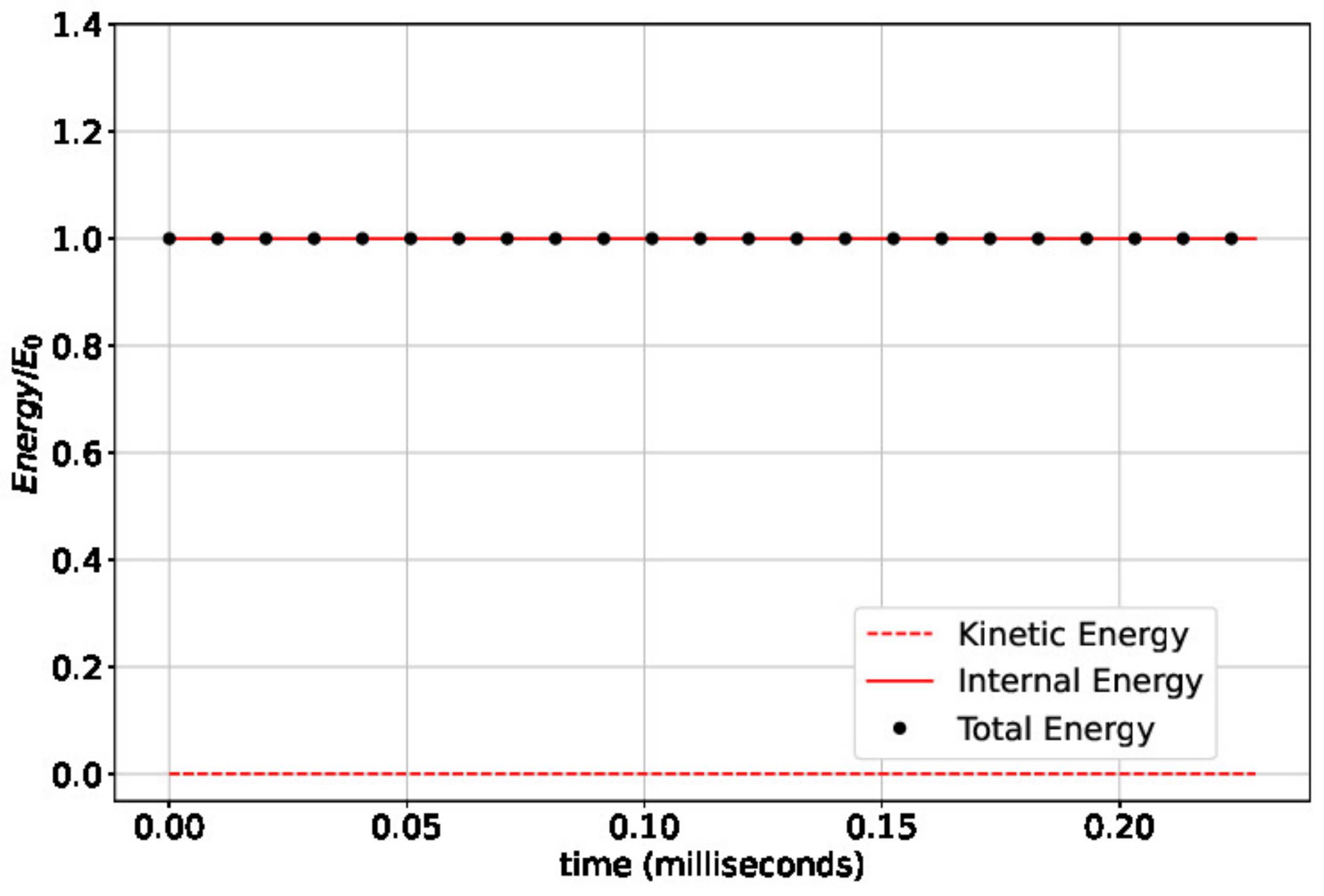}
\caption{Simulation of hydrogen/air constant volume perfectly stirred reactor. Top: History of temperature and hydroxide mass fraction. Bottom: History of the kinetic, the internal and the total energy. All quantities are scaled by the initial total energy $E_0$. }
\label{fig:psr}
\end{figure}

A constant volume PSR is simulated using LBM with a three-dimensional domain consisting of $4 \times 4 \times 4$ nodes. Periodic boundary conditions are used in all directions. The computational domain is initialized with a stagnant and homogeneous hydrogen/air mixture at an equivalence ratio $\phi=1$, pressure $P_{\rm in}=1\, \text{atm}$ and temperature $T_{\rm in}=1400\, \text{K}$. Fig.\ \ref{fig:psr} shows the evolution of the temperature and of the hydroxide mass fraction in the reactor over time. The results from the lattice Boltzmann model are compared to the solution produced by the ideal gas constant volume reactor from Cantera. The time integration in Cantera is performed through its built-in `advance' function. Accurate match with the results obtained from Cantera verifies that the coupling and the multistep time integration of the reaction term is correct. Since all the boundaries are periodic in this setup, the total energy of the system must remain constant. Also, due to the completely homogeneous initial condition, no kinetic energy should develop over time. Fig.\ \ref{fig:psr} verifies that in the absence of flow, the total energy not only equals the internal energy but it also remains constant in time, as expected. 

\subsection{Laminar flame speed}
\label{sec:flamespeed}

\begin{figure}
	\centering
		\includegraphics[width=0.85\linewidth]{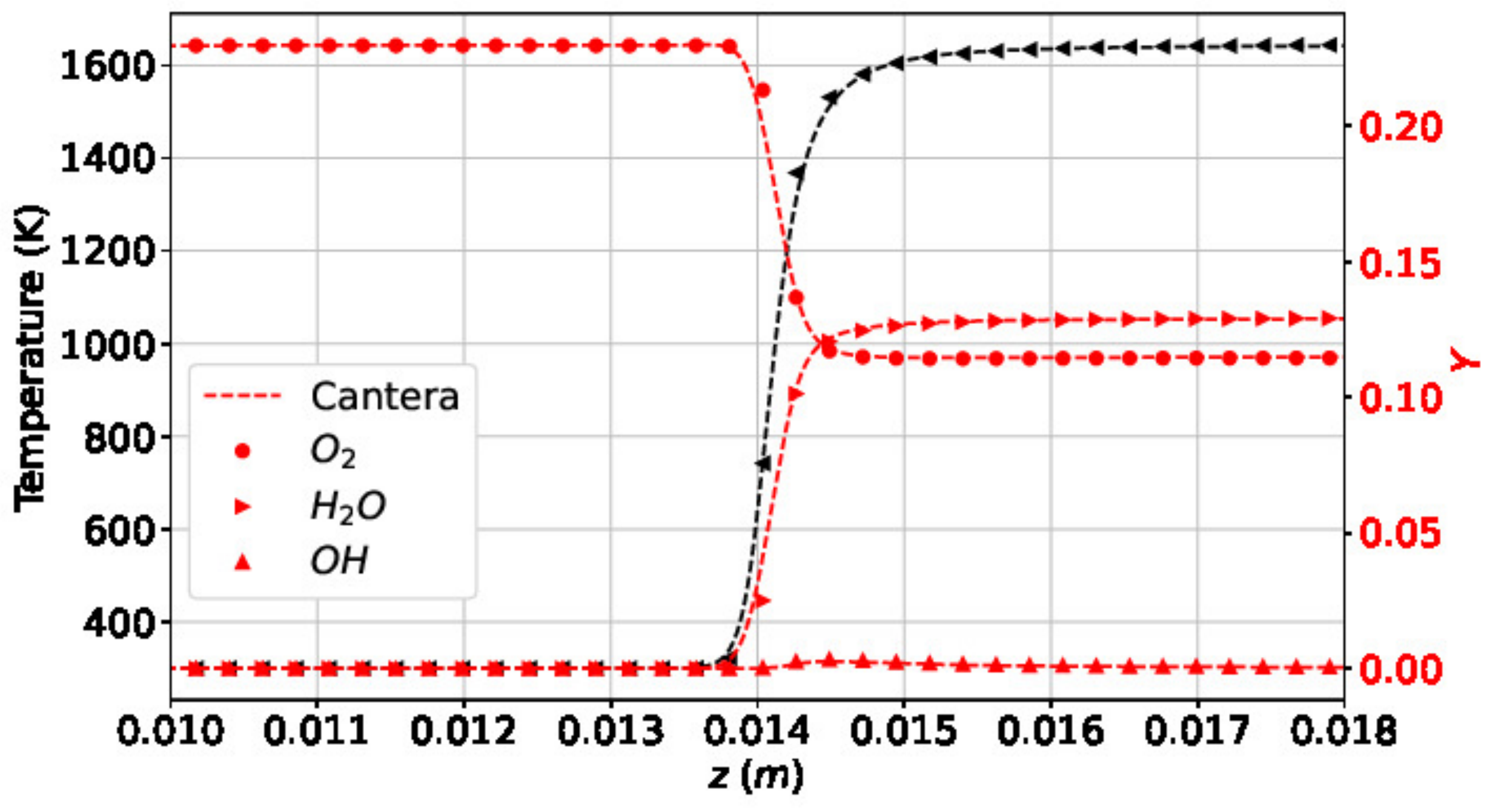}
		\caption{Profiles of temperature and mass fractions for 1D planar flame at $\phi=0.5$.}
		\label{fig:flameProfile}
\end{figure}

For a further validation, we calculate the burning velocity of a hydrogen/air mixture. The setup 
consists of a one-dimensional tube initialized with unburnt mixture at $T_{\rm u}=300\, \text{K}$ throughout from the left end up to $80\%$ of the domain towards the right. The remaining $20\%$ of the domain are initialized with the adiabatic flame temperature $T_{\rm af}=1642.49\, \text{K}$ and with the equilibrium burnt composition at the respective equivalence ratio. The pressure is initialized uniformly at $P_{\rm in}=1\, \text{atm}$. The inlet and the outlet boundary condition used in this case will be explained below in sec.\ \ref{sec:bc}. 
At the left end, the inlet velocity is set to $u_{\rm in}=10\, \text{cm/s}$ so that the flame propagates from right to left against the unburnt mixture. 

We use the laminar flame thickness $\delta_{\rm f}=0.043\, \text{cm}$ at $\phi=0.5$ for defining the reference length, 
where $\delta_{\rm f}=(T_{\rm af}-T_{\rm u})/\max(\left|{dT}/{dx}\right|)$.
The domain size is $N \approx 23 \delta_{\rm f}$ with a resolution of $34$ points per flame thickness.  As evident in Fig. \ref{fig:flameProfile}, the profiles of the temperature and the mass fractions for $\phi=0.5$ compare well with the solution obtained from the `FreeFlame' solver of Cantera. 
The burning velocity is found to be $S_L=59\, \text{cm/s}$ which is in good agreement with the reference result of  \cite{pizza_dynamics_2008}, i.e. $58\, \text{cm/s}$. 
To summarize, the basic validation of the proposed LBM  for reactive mixtures is considered successful. We now proceed with specifying various boundary conditions for the multicomponent LBM, needed for most of practical applications.
\section{Boundary Conditions}
\label{sec:bc}
\subsection{Nomenclature}
\begin{figure}
     \centering
     \includegraphics[width=0.8\linewidth]{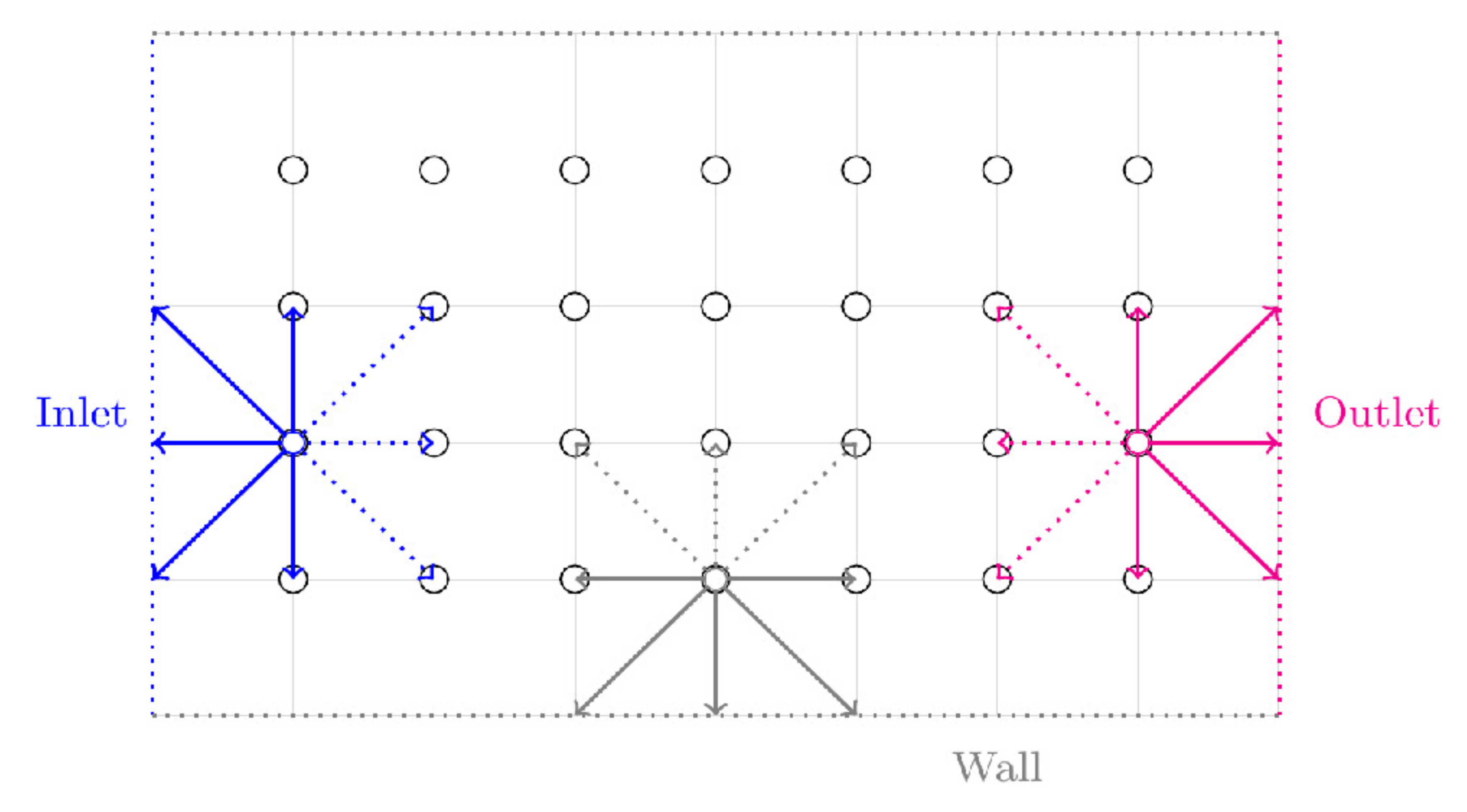}
      \caption{Schematic of the LBM computational domain. Empty circles represent computational nodes and coloured dotted edges indicate boundaries. At interface nodes, dotted arrows are the incoming velocities $\bm{c}_i$, $i\in\mathcal{D}$ while solid arrows are the outgoing velocities $\bm{c}_i$, $i\notin\mathcal{D}$.}
     \label{fig:bc}
\end{figure}

Boundary conditions for multi-component LBM are scarce in the literature.
In order to facilitate the explanation, we use the cartoon in Fig.\ \ref{fig:bc}, which represents a rectangular grid and empty circles represent grid points (nodes) which are part of the computational domain. The boundaries are marked by coloured dotted lines, where the colour reflects either the wall, the inlet or the outlet. The boundaries do not belong to the computational domain and therefore do not participate in the collision and the advection operations. During the advection step, a node at location $\bm{x}$ performs the following operation for each of the populations $f_i$,
\begin{equation}
  f_i(\bm{x},t) = f_i(\bm{x}-\bm{c}_i \delta t,t-\delta t).
\label{eqn:advection}
\end{equation}
Equation (\ref{eqn:advection}) is a mathematical expression for the free streaming of a population $f_i$ by jumping a distance $\bm{c}_i \delta t$ to a new node. Since we do not need to discuss the collision step in this section, the times $t$ and $t-\delta t$ simply indicate the post- and the pre-advection states, respectively. In Fig.\ \ref{fig:bc}, each population $f_i$ is represented by its corresponding discrete velocity vector $\bm{c}_i$ (link) by an arrow pointing in the direction of its propagation. Solid arrows represent the post-advection populations that arrived from a node belonging to the computational domain. Dotted arrows represent the post-advection populations that have arrived from one of the boundaries and carry with them the information about the fluid properties at the boundary. These populations will be referred to as incoming populations since they enter the domain from the boundaries. In the lattice Boltzmann method, the boundary conditions are applied by specifying the incoming populations. The nodes which are adjacent to the boundaries and therefore require such description for incoming populations will be referred to as the interface nodes. We denote $\mathcal{D}$ the set of the incoming velocities at the interface node. 
Finally, the rest of the velocities at the interface node $\bm{c}_i\notin \mathcal{D}$ are the outgoing velocities.

Below, the equilibrium form shall be used to evaluate a variety of incoming populations. 
In order to keep the discussion concise, we shall display the dependence of pertinent equilibria on the respective control parameters as follows:
	\begin{align}
		&f_{ai}^{\rm eq}=f_{ai}^{\rm eq}\left(\rho Y_a,\bm{u},T\right),\label{eq:faEqGeneral}\\
		&f_{i}^{\rm eq}=f_{i}^{\rm eq}\left(\rho,Y,\bm{u},T\right),\label{eq:fEqGeneral}\\
		&g_{i}^{\rm eq}=g_{i}^{\rm eq}\left(\rho,Y,\bm{u},T\right).\label{eq:gEqGeneral}
	\end{align}
Here $Y=\left\{Y_1,\dots, Y_M\right\}$ stands for the totality of mass fractions.
Dependence on the mixture composition $Y$ in the energy lattice equilibrium (\ref{eq:gEqGeneral}) is manifest in the operational definition (\ref{eq:geq_i}) through the mixture-averaged gas constant $R$ in the operators (\ref{eqn:oalpha}) as well as in the mixture energy (\ref{eq:totalE}). The composition dependence $Y$ enters the momentum lattice equilibrium (\ref{eq:fEqGeneral}) through the gas constant $R$, cf.\ Eqs.\ (\ref{eqn:Peq}) and (\ref{eq:feq}).
We now proceed to derive the wall, the inlet and the outlet boundary conditions for the multicomponent LBM. 

\subsection{Wall boundary conditions}
\label{sec:wallbc}
\subsubsection{Bounce-back boundary condition}
Bounce-back (BB) is a widely used wall boundary condition in the lattice Boltzmann method \citep{ladd_numerical_1994}. 
For the incoming populations at interface node $\bm{x}$, the bounce-back rule reads,
\begin{equation}
    {f_i^{\rm bb}(\bm{x},t) = f_k(\bm{x},t-\delta t), \bm{c}_k=-\bm{c}_i, \ \text{if } i \in \mathcal{D}}.
    \label{eqn:bb}
\end{equation}
Here $\mathcal{D}$ is the set of incoming velocities shown by grey dotted arrows in Fig.\ \ref{fig:bc}.
When applied on the momentum lattice, the bounce-back rule (\ref{eqn:bb}) results in the no-slip boundary condition at a half-way distance between the wall and the interface nodes \citep{ziegler_boundary_1993,chen_lattice_1998,boyd_application_2004}.
On the energy lattice, the bounce-back boundary condition conserves the total energy and leads to zero heat flux, thereby representing an adiabatic wall \citep{he_novel_1998}. 
While simple and efficient, the bounce-back boundary condition (\ref{eqn:bb}) is limited as it does not allow to impose a target value for the velocity at a prescribed wall location nor to implement a target wall temperature. 
Since these are the cases typical of many applications, including the ones considered below, a so-called 
Tamm--Mott-Smith (TMS) boundary condition of \citet{chikatamarla_entropic_2013} shall be adapted to the multicomponent mixture. 

\subsubsection{Tamm--Mott-Smith wall boundary condition}
\label{sec:tms}

Let $\bm{u}^{\rm tgt}$, $\bm{u}_a^{\rm tgt}$ and $T^{\rm tgt}$ be the target values of the flow velocity, species velocity and the temperature, respectively, to be imposed at the interface node $\bm{x}$. 
Moreover, the outgoing populations $f_i$, $g_i$ and $f_{ai}$, where $i\notin \mathcal{D}$, are obtained in the propagation step (\ref{eqn:advection}) and assumed known.	
	The TMS construction of the incoming populations $f_i^{\rm TMS}$, $g_i^{\rm TMS}$ and $f_{ai}^{\rm TMS}$, where $i\in \mathcal{D}$, executes the following steps:
\begin{enumerate}
    \item \noindent Perform bounce-back on the momentum lattice to find the densities $\rho^{\rm bb}$,
    \begin{align}
    	{\rho^{\rm bb}=\sum_{i\in \mathcal{D}}f_i^{\rm bb}+\sum_{i\notin \mathcal{D}}f_i}.\label{eq:BBrho}
    \end{align}
     Perform bounce-back on the species lattices to find the mass fractions ${Y}^{\rm bb}$,
    \begin{align}
    	{\rho^{\rm bb}{Y}_a^{\rm bb}=\sum_{i\in \mathcal{D}}f_{ai}^{\rm bb}+\sum_{i\notin \mathcal{D}}f_{ai},\ a=1,\dots,M}.\label{eq:BBY}
    \end{align}
Note that the bounce-back operation is used solely for computing the density $\rho^{\rm bb}$ (\ref{eq:BBrho}) and the mass fractions $Y^{\rm bb}$ (\ref{eq:BBY}), in order to satisfy mass conservation at the boundary. However, the incoming populations are {\it not} set to the bounce-back values $f_i^{\rm bb}$, rather, they are defined with the subsequent steps of the TMS algorithm.
    \item \label{TMS:eq}\noindent  The bounce-back density  $\rho^{\rm bb}$ (\ref{eq:BBrho}) and mass fractions ${Y}^{\rm bb}$ (\ref{eq:BBY}), together with the target velocities $\bm{u}^{\rm tgt}$, 
    $\bm{u}_a^{\rm tgt}$ 
    and temperature $T^{\rm tgt}$, uniquely specify the equilibrium states  $f_i^{\rm tgt}$,  $g_i^{\rm tgt}$ and $f_{ai}^{\rm tgt}$ at the interface node,
        \begin{align}
        	&f_i^{\rm tgt}= f_i^{\rm eq}(\rho^{\rm bb},Y^{\rm bb},\bm{u}^{\rm tgt},T^{\rm tgt}), \label{eq:tgtf}\\
        	&g_i^{\rm tgt}= g_i^{\rm eq}(\rho^{\rm bb},Y^{\rm bb},\bm{u}^{\rm tgt},T^{\rm tgt}),\label{eq:tgtg}\\
        	&f_{ai}^{\rm tgt}= f_{ai}^{\rm eq}(\rho^{\rm bb}Y_a^{\rm bb},\bm{u}_a^{\rm tgt},T^{\rm tgt}).\label{eq:tgtfa}
        \end{align}
    \label{enum:wallDist}
 \item \label{TMS:loc}\noindent With the incoming populations set to the target equilibrium, we find the local density $\rho^{\rm loc}$, flow velocity $\bm{u}^{\rm loc}$, mass fractions ${Y}^{\rm loc}$, species velocities $\bm{u}_a^{\rm loc}$ and temperature $T^{\rm loc}$ at the interface node,
    \begin{align}
    	&\sum_{i\in\mathcal{D}} f_i^{\rm tgt}  +\sum_{i\notin\mathcal{D}} f_i =  \rho^{\rm loc},  \label{eq:locRho}\\
    	&\sum_{i\in\mathcal{D}} \bm{c}_if_i^{\rm tgt}  +\sum_{i\notin\mathcal{D}} \bm{c}_if_i =  \rho^{\rm loc} \bm{u}^{\rm loc}, \label{eq:locu} \\
    	&{\sum_{i\in \mathcal{D}}f_{ai}^{\rm tgt}+\sum_{i\notin \mathcal{D}}f_{ai}=\rho^{\rm loc}{Y}_a^{\rm loc},\ a=1,\dots,M}. \label{eq:locY}  \\
&{\sum_{i\in \mathcal{D}} \bm{c}_i f_{ai}^{\rm tgt}+\sum_{i\notin \mathcal{D}} \bm{c}_i f_{ai}=\rho_a^{\rm loc}\bm{u}_a^{\rm loc},\ a=1,\dots,M}. \label{eq:locua}  \\
    	&\sum_{i\in\mathcal{D}} g_i^{\rm tgt}  +\sum_{i\notin\mathcal{D}} g_i =  
    	\rho^{\rm loc}\sum_{a=1}^M{Y^{\rm loc}_a}\left[U_a^0 + \int_{T_0}^{T^{\rm loc}}  {C}_{a,v}(T')dT'\right]+\frac{1}{2}\rho^{\rm loc}\left(u^{\rm loc}\right)^2.\label{eq:locT}
    \end{align}
    We remind that Eq.\ (\ref{eq:locT}) is an integral equation to be resolved for the temperature $T^{\rm loc}$, cf.\ sec.\ \ref{sec:energy}, Eq.\ (\ref{eq:temperature}). 
    With the local parameters, the following equilibrium populations are uniquely specified on the momentum, energy and species lattices,
     \begin{align}
    	&f_i^{\rm loc}= f_i^{\rm eq}(\rho^{\rm loc},Y^{\rm loc},\bm{u}^{\rm loc},T^{\rm loc}), \\
    	&g_i^{\rm loc}= g_i^{\rm eq}(\rho^{\rm loc},Y^{\rm loc},\bm{u}^{\rm loc},T^{\rm loc}), \\
    	&{f_{ai}^{\rm loc}= f_{ai}^{\rm eq}(\rho^{\rm loc}Y_a^{\rm loc},\bm{u}_a^{\rm loc},T^{\rm loc}).}
    \end{align}
    
    \item \label{TMS:fin} \noindent Finally, we update all populations of the momentum, the energy and the species lattices at the wall interface node as:
    \begin{align}
    &f^{\rm TMS}_i = \left\{\begin{aligned}&2f_i^{\rm tgt} - f_i^{\rm loc}, &\text{ if } i \in \mathcal{D},& \\
    	& f_i^{\rm tgt}+f_i - f_i^{\rm loc}, & \text{ if } i \notin \mathcal{D}&\\
    	\end{aligned}\right. \label{eqn:ftms}\\
     &g^{\rm TMS}_i = \left\{\begin{aligned}&2g_i^{\rm tgt} - g_i^{\rm loc}, &\text{ if } i \in \mathcal{D},& \\
    	&g_i^{\rm tgt} + g_i - g_i^{\rm loc}, & \text{ if } i \notin \mathcal{D}&\\
    \end{aligned}\right. \label{eqn:gtms}\\
    &{f^{\rm TMS}_{ai}} = \left\{\begin{aligned}&2f_{ai}^{\rm tgt} - f_{ai}^{\rm loc}, &\text{ if } i \in \mathcal{D},& \\
    	& f_{ai}^{\rm tgt}+f_{ai} - f_{ai}^{\rm loc}, & \text{ if } i \notin \mathcal{D}&\\
    	\end{aligned}\right. \label{eqn:fatms}
    \end{align}
\end{enumerate}
Comments are in order. The TMS boundary condition in step (\ref{TMS:fin}) sets  the flow variables at the interface nodes to 
$\rho^{\rm bb}$, ${Y}^{\rm bb}$, $\bm{u}_a^{\rm tgt}$, $\bm{u}^{\rm tgt}$
and $T^{\rm tgt}$. While the same is achieved by the target equilibrium at step (\ref{TMS:eq}), the corresponding equilibrium boundary condition is insufficient as it is prone to generating spurious shocks, cf.\ \citet{chikatamarla_entropic_2013}. For this reason, a non-equilibrium part of the incoming populations is taken into consideration and modelled with the local state in step (\ref{TMS:loc}). Note that, while the latter also uses the equilibrium form, it is evaluated at different (local) values of flow variables and thus describes a non-equilibrium state relative to the target equilibrium. The presence of two different equilibrium states in the resulting populations motivated \cite{chikatamarla_entropic_2013} to naming the algorithm in analogy to the bimodal Tamm--Mott-Smith shock wave approximation for the Boltzmann equation \citep{mott-smith_solution_1951}. 

With the exception of walls aligned with the Cartesian LBM grid, target parameters $\bm{u}^{\rm tgt}$, $\bm{u}_a^{\rm tgt}$ and $T^{\rm tgt}$ at the interface nodes are obtained by interpolation between the values of the corresponding fields at the wall, $\bm{u}^{\rm wall}$, $\bm{u}_a^{\rm wall}$ and $T^{\rm wall}$, and the data at the surrounding fluid nodes. Interpolation is performed following the procedure described in \citet{chikatamarla_entropic_2013,dorschner_grads_2015}, examples shall be demonstrated below in sec.\ \ref{sec:microtube} for stationary no-slip walls,  $\bm{u}^{\rm wall}=0$, subject to a temperature profile. Finally, the impermeable wall boundary condition is imposed on the species populations by setting the species velocity at the wall as $\bm{u}_a^{\rm wall}=\bm{u}^{\rm wall}$. 
Zero flux of species at the wall is implied by the absence of diffusion velocity in the species equilibrium velocity $\bm{u}^{\rm wall}$.

\subsection{Inlet}
\label{sec:inletbc}
The flux boundary condition is widely used to model the inlet in  multicomponent flows \citep{kee_chemically_2003,pizza_three-dimensional_2010,pizza_dynamics_2008,goodwin_cantera_2018}. 
The rationale of this boundary condition is that it prescribes only the incoming mass fluxes of species $\rho_a^{\rm in}\bm{u}^{\rm in}$. Because only the incoming mass flux is prescribed and not the mass itself, the composition at the inlet interface node is not fixed to the incoming composition $Y^{\rm in}$. This degree of freedom is necessary as light species such as hydrogen have the capability to diffuse fast enough and thus are able to propagate upstream into the inlet. 
Therefore, the composition at the inlet interface node is not a  fixed set of parameters but is rather a result of a balance between the mass flux inside the domain and the inlet mass flux. Below, we establish the flux boundary condition for the multicomponent lattice Boltzmann setting.

In Fig.\ \ref{fig:bc}, the inlet boundary is represented by a dotted vertical blue line. The inlet boundary condition is applied on the interface nodes where incoming discrete velocities $\bm{c}_i$, $i\in\mathcal{D}$, are represented by dotted blue arrows. 
With the inlet data for  mass flux $\rho^{\rm in} \bm{u}^{\rm in}$, composition ${Y}^{\rm in}$ and temperature $T^{\rm in}$, populations at the interface node are derived in the following steps:

\begin{enumerate}
    \item \noindent The inlet density  $\rho^{\rm in}$ and composition ${Y}^{\rm in}$, together with the inlet velocity $\bm{u}^{\rm in}$ and temperature $T^{\rm in}$, uniquely specify the inlet equilibrium populations  $f_i^{\rm in}$,  $g_i^{\rm in}$ and $f_{ai}^{\rm in}$ at the inlet interface node,
            \begin{align}
        	&f_i^{\rm in}= f_i^{\rm eq}(\rho^{\rm in},Y^{\rm in},\bm{u}^{\rm in},T^{\rm in}), \label{eq:in_f}\\
        	&g_i^{\rm in}= g_i^{\rm eq}(\rho^{\rm in},Y^{\rm in},\bm{u}^{\rm in},T^{\rm in}),\label{eq:in_g}\\
        	&f_{ai}^{\rm in}= f_{ai}^{\rm eq}(\rho^{\rm in}Y_a^{\rm in},\bm{u}^{\rm in},T^{\rm in}).\label{eq:in_fa}
        \end{align}
    \item \label{inlet:loc}\noindent With the incoming populations set to the inlet equilibrium and the outgoing populations known, we find the local density $\rho^{\rm loc}$, flow velocity $\bm{u}^{\rm loc}$, composition ${Y}^{\rm loc}$
    and temperature $T^{\rm loc}$ at the interface node,    
    \begin{align}
&\sum_{i\in\mathcal{D}} f_i^{\rm in}  +\sum_{i\notin\mathcal{D}} f_i =  \rho^{\rm loc},  \label{eq:inletlocRho}\\
&\sum_{i\in\mathcal{D}} \bm{c}_if_i^{\rm in}  +\sum_{i\notin\mathcal{D}} \bm{c}_if_i =  \rho^{\rm loc} \bm{u}^{\rm loc}, \label{eq:inletlocu} \\
&{\sum_{i\in \mathcal{D}}f_{ai}^{\rm in}+\sum_{i\notin \mathcal{D}}f_{ai}=\rho^{\rm loc}{Y}_a^{\rm loc},\ a=1,\dots,M}. \label{eq:inletlocY}  \\
&\sum_{i\in\mathcal{D}} g_i^{\rm in}  +\sum_{i\notin\mathcal{D}} g_i =  
\rho^{\rm loc}\sum_{a=1}^M{Y^{\rm loc}_a}\left[U_a^0 + \int_{T_0}^{T^{\rm loc}}  {C}_{a,v}(T')dT'\right]+\frac{1}{2}\rho^{\rm loc}\left(u^{\rm loc}\right)^2.\label{eq:inletlocT}
    \end{align}
    With the local parameters, the following equilibrium populations are uniquely specified on the momentum, energy and species lattices,
     \begin{align}
    	&f_i^{\rm loc}= f_i^{\rm eq}(\rho^{\rm loc},Y^{\rm loc},\bm{u}^{\rm loc},T^{\rm loc}), \label{eq:in_loc_f}\\
    	&g_i^{\rm loc}= g_i^{\rm eq}(\rho^{\rm loc},Y^{\rm loc},\bm{u}^{\rm loc},T^{\rm loc}), \label{eq:in_loc_g}\\
    	&{f_{ai}^{\rm loc}= f_{ai}^{\rm eq}(\rho^{\rm loc}Y_a^{\rm loc},\bm{u}^{\rm loc},T^{\rm loc}).\label{eq:in_loc_fa}}
    \end{align}
\item \noindent Replacing the local flow velocity $\bm{u}^{\rm loc}$ and temperature $T^{\rm loc}$ with the target values $\bm{u}^{\rm in}$ and $T^{\rm in}$, the following target equilibrium populations are identified, 
            \begin{align}
        	&f_i^{\rm tgt}= f_i^{\rm eq}(\rho^{\rm loc},Y^{\rm loc},\bm{u}^{\rm in},T^{\rm in}), \label{eq:in_tgt_f}\\
        	&g_i^{\rm tgt}= g_i^{\rm eq}(\rho^{\rm loc},Y^{\rm loc},\bm{u}^{\rm in},T^{\rm in}),\label{eq:in_tgt_g}\\
        	&f_{ai}^{\rm tgt}= f_{ai}^{\rm eq}(\rho^{\rm loc}Y_a^{\rm loc},\bm{u}^{\rm in},T^{\rm in}).\label{eq:in_tgt_fa}
        \end{align}
\item  \noindent Finally, all populations at the inlet interface nodes are updated as follows:
    \begin{align}
    &f^{\rm inlet}_i = \left\{\begin{aligned}&f_i^{\rm tgt} + f_i^{\rm in} - f_i^{\rm loc}, &\text{ if } i \in \mathcal{D},& \\
    	& f_i^{\rm tgt}+f_i - f_i^{\rm loc}, & \text{ if } i \notin \mathcal{D}&\\
    	\end{aligned}\right. \label{eqn:f_inlet}\\
     &g^{\rm inlet}_i = \left\{\begin{aligned}&g_i^{\rm tgt} + g_i^{\rm in} - g_i^{\rm loc}, &\text{ if } i \in \mathcal{D},& \\
    	&g_i^{\rm tgt} + g_i - g_i^{\rm loc}, & \text{ if } i \notin \mathcal{D}&\\
    \end{aligned}\right. \label{eqn:g_inlet}\\
    &{f^{\rm inlet}_{ai}} = \left\{\begin{aligned}&f_{ai}^{\rm tgt} + f_{ai}^{\rm in} - f_{ai}^{\rm loc}, &\text{ if } i \in \mathcal{D},& \\
    	& f_{ai}^{\rm tgt}+f_{ai} - f_{ai}^{\rm loc}, & \text{ if } i \notin \mathcal{D}&\\
    	\end{aligned}\right. \label{eqn:fa_inlet}
    \end{align}
     \label{step:applyInlet}
\end{enumerate}

It is straightforward to verify that the populations (\ref{eqn:f_inlet}), (\ref{eqn:g_inlet}) and (\ref{eqn:fa_inlet}) at step (\ref{step:applyInlet}) imply 
the target values $\bm{u}^{\rm in}$ and $T^{\rm in}$ for the velocity and temperature at the interface node, respectively. At the same time, the composition and density at the interface node are identified as  ${Y}^{\rm loc}$ and $\rho^{\rm loc}$, respectively. The latter  are derived in step (\ref{inlet:loc}) by taking into account the outgoing populations and are different, in general, from the inlet values 
	${Y}^{\rm in}$ and $\rho^{\rm in}$. Thus, the outgoing populations contribute to the balance between the incoming and outgoing mass fluxes  as required by the flux boundary condition. It is instructive to compare with the TMS wall boundary condition of sec.\ \ref{sec:tms} where the local composition at the interface node was determined by the bounce-back step, Eqs.\ (\ref{eq:BBrho}) and (\ref{eq:BBY}). In the present case, at step (\ref{inlet:loc}), the local composition is computed using the equilibrium at the inflow properties for the incoming populations.  
Note that, while the inlet composition $Y^{\rm loc}$ is already determined at step (\ref{inlet:loc}), the purpose of the remaining steps is to enforce the inlet velocity $\bm{u}^{\rm in}$ and temperature $T^{\rm in}$. Hence, whenever the inlet temperature and velocity need not be strictly imposed, it is sufficient to terminate the algorithm at step (\ref{inlet:loc}) and to apply the local equilibria (\ref{eq:in_loc_f}), (\ref{eq:in_loc_g}) and (\ref{eq:in_loc_fa}). With this simplification, the velocity and temperature acquire local values $\bm{u}^{\rm loc}$ and $T^{\rm loc}$, respectively, rather than the target values $\bm{u}^{\rm in}$ and $T^{\rm in}$.
The latter simplification was validated in \citet{sawant_consistent_2021} with the simulation of diffusion in opposed jets. The mass fractions at the inlets of both jets matched the reference solution  by Cantera, which employs a macroscopic realization of the flux boundary condition. 
While the simplified inlet realization (\ref{eq:in_loc_f}), (\ref{eq:in_loc_g}) and (\ref{eq:in_loc_fa}) can be regarded as a good approximation to the macroscopic flux boundary condition,  
in this work we rather use the inlet populations (\ref{eqn:f_inlet}), (\ref{eqn:g_inlet}) and (\ref{eqn:fa_inlet}) to ensure that the inlet velocity and temperature are imposed exactly.

\subsection{Outlet}
\label{sec:outlet}
Unlike the inlet and the wall, the values of the macroscopic state variables are usually unknown at the outlet. To that end, we apply the  Local One Dimensional Inviscid (LODI) approximation by \citet{poinsot_boundary_1992}. LODI is based on the characteristics of compressible Euler equations, i.e. Eqs.\ (\ref{eqn:dtrho}), (\ref{eqn:dtu}) and (\ref{eqn:dtE}) without dissipation terms. LODI boundary condition allows both the pressure fluctuations travelling as sound waves as well as the convection disturbances travelling as entropy waves  to exit the computational domain with minimum reflection \citep{poinsot_boundary_1992}. The LODI approximation is derived for a single-component fluid and therefore predicts the outlet density $\rho^{\rm out}$, velocity $\bm u^{\rm out}$ and the temperature $T^{\rm out}$ which can be directly used in the present mean field formulation of the mixture. In addition, we need also to specify the composition ${Y}^{\rm out}$ at the outlet. Consistent with the LODI approximation, we use the advection part of the species equation (\ref{eq:dtrhoa}) which is discretized at the outlet interface node with forward Euler scheme to give,
\begin{equation}
    Y_a^{\rm out}(\bm{x},t)  = Y_a^{\rm out}(\bm{x},t- \delta t) -\delta t \bm{u}^{\rm out}(\bm{x},t- \delta t) \cdot\nabla Y_a^{\rm out}(\bm x,t- \delta t),
    \label{eqn:convectiveBC}
\end{equation}
where mass fraction $Y_a^{\rm out}(\bm{x},t- \delta t)$ is known from the previous time step, while $\bm{u}^{\rm out}$ is the LODI outlet velocity. The gradient $\nabla Y_a^{\rm out}$ is evaluated by backward finite difference. 
Armed with the outlet data, we proceed to specify the populations at the outlet interface node, following essentially the steps already familiar from the wall and inlet construction:
\begin{enumerate}
    \item \label{outlet:out} \noindent Outlet data $Y_a^{\rm out}$, $\rho^{\rm out},\bm{u}^{\rm out}$ and $T^{\rm out}$ uniquely specifies the equilibrium populations  $f_i^{\rm out}$,  $g_i^{\rm out}$ and $f_{ai}^{\rm out}$ at the outlet interface node,
            \begin{align}
        	&f_i^{\rm out}= f_i^{\rm eq}(\rho^{\rm out},Y^{\rm out},\bm{u}^{\rm out},T^{\rm out}), \label{eq:out_f}\\
        	&g_i^{\rm out}= g_i^{\rm eq}(\rho^{\rm out},Y^{\rm out},\bm{u}^{\rm out},T^{\rm out}),\label{eq:out_g}\\
        	&f_{ai}^{\rm out}= f_{ai}^{\rm eq}(\rho^{\rm out} Y_a^{\rm out},\bm{u}^{\rm out},T^{\rm out}).\label{eq:out_fa}
        \end{align}
    \item \label{outlet:eq}\noindent With the incoming populations set to the outlet equilibrium, we find the local density $\rho^{\rm loc}$, flow velocity $\bm{u}^{\rm loc}$, mass fractions ${Y}^{\rm loc}$,
    and temperature $T^{\rm loc}$ at the outlet interface node,    
    \begin{align}
&\sum_{i\in\mathcal{D}} f_i^{\rm out}  +\sum_{i\notin\mathcal{D}} f_i =  \rho^{\rm loc},  \label{eq:outletlocRho}\\
&\sum_{i\in\mathcal{D}} \bm{c}_i f_i^{\rm out}  +\sum_{i\notin\mathcal{D}} \bm{c}_if_i =  \rho^{\rm loc} \bm{u}^{\rm loc}, \label{eq:outletlocu} \\
&{\sum_{i\in \mathcal{D}}f_{ai}^{\rm out}+\sum_{i\notin \mathcal{D}}f_{ai}=\rho^{\rm loc}{Y}_a^{\rm loc},\ a=1,\dots,M}. \label{eq:outletlocY}  \\
&\sum_{i\in\mathcal{D}} g_i^{\rm out}  +\sum_{i\notin\mathcal{D}} g_i =  
\rho^{\rm loc}\sum_{a=1}^M{Y^{\rm loc}_a}\left[U_a^0 + \int_{T_0}^{T^{\rm loc}}  {C}_{a,v}(T')dT'\right]+\frac{1}{2}\rho^{\rm loc}\left(u^{\rm loc}\right)^2.\label{eq:outletlocT}
    \end{align}
Based on these local parameters, the local equilibrium populations are uniquely specified on the momentum, energy and species lattices,
     \begin{align}
    	&f_i^{\rm loc}= f_i^{\rm eq}(\rho^{\rm loc},Y^{\rm loc},\bm{u}^{\rm loc},T^{\rm loc}), \\
    	&g_i^{\rm loc}= g_i^{\rm eq}(\rho^{\rm loc},Y^{\rm loc},\bm{u}^{\rm loc},T^{\rm loc}), \\
    	&{f_{ai}^{\rm loc}= f_{ai}^{\rm eq}(\rho^{\rm loc}Y_a^{\rm loc},\bm{u}^{\rm loc},T^{\rm loc}).}
    \end{align}
\item  \noindent Finally, all populations of the momentum, energy and species lattices at the outlet interface node are updated as:
    \begin{align}
    &f^{\rm outlet}_i = \left\{\begin{aligned}& 2 f_i^{\rm out} - f_i^{\rm loc}, &\text{ if } i \in \mathcal{D},& \\
    	& f_i^{\rm out}+f_i - f_i^{\rm loc}, & \text{ if } i \notin \mathcal{D}&\\
    	\end{aligned}\right. \label{eqn:f_outlet}\\
     &g^{\rm outlet}_i = \left\{\begin{aligned}& 2 g_i^{\rm out} - g_i^{\rm loc}, &\text{ if } i \in \mathcal{D},& \\
    	&g_i^{\rm out} + g_i - g_i^{\rm loc}, & \text{ if } i \notin \mathcal{D}&\\
    \end{aligned}\right. \label{eqn:g_outlet}\\
    &{f^{\rm outlet}_{ai}} = \left\{\begin{aligned}& 2 f_{ai}^{\rm out} - f_{ai}^{\rm loc}, &\text{ if } i \in \mathcal{D},& \\
    	& f_{ai}^{\rm out}+f_{ai} - f_{ai}^{\rm loc}, & \text{ if } i \notin \mathcal{D}&\\
    	\end{aligned}\right. \label{eqn:fa_outlet}
    \end{align}
     \label{step:applyOutlet}
\end{enumerate}

With populations (\ref{eqn:f_outlet}), (\ref{eqn:g_outlet}) and (\ref{eqn:fa_outlet}) at step (\ref{step:applyOutlet}), the macroscopic fields at the outlet are set to the target values $\rho^{\rm out},\bm{u}^{\rm out}$, $T^{\rm out}$ and $Y^{\rm out}$, 
as prescribed by LODI approximation and (\ref{eqn:convectiveBC}). 
Although the same is achieved by the equilibrium populations at step (\ref{outlet:out}), the non-equilibrium part of the incoming populations is taken into consideration and modelled with the local state in step (\ref{outlet:eq}). Thus, the present construction of the outlet is similar to the TMS wall boundary condition of sec.\ \ref{sec:tms}.  

\section{Wall-bounded reactive flow}
\label{sec:results}

In order to test the proposed boundary conditions, we perform the computation of combustion in microtubes. 
The results are validated with the direct numerical simulation of \citet{pizza_dynamics_2008-1} for $2D$ microchannels and with \citet{pizza_three-dimensional_2010} for a $3D$ microtube. The setup involves combustion of a premixed hydrogen/air mixture in a tube over a range of inlet velocities $u_{in}$ of the unburnt mixture. The fuel-lean unburnt mixture of equivalence ratio $\phi=0.5$ at a temperature $T_u=300$ K and pressure $1 \;\text{atm}$ enters a microchannel with $l/d=10$. Here, $l$ is the length of the tube and $d$ is its diameter. The mixture gets ignited due to hot isothermal walls which are maintained at a temperature of $T_w=960\, \text{K}$. The wall temperature is increased from $300\, \text{K}$  at the inlet to $960\, \text{K}$  using a hyperbolic tangent profile at a distance of about $l/20$ from the inlet. 
For this premixed initial condition, the burning velocity is obtained as $S_L=59\, \text{cm/s}$ and the flame thickness is obtained as $\delta_{\rm f}=0.043\, \text{cm}$ from solving a $1D$ flame propagation setup with the lattice Boltzmann method in section (\ref{sec:flamespeed}).

\subsection{Premixed hydrogen/air flames in a microchannel} \label{sec:microchannel}

For the $2D$ simulations, we choose a channel diameter $d=1\, \text{mm}$, which corresponds to a width of $2.3256 \;\delta_{\rm f}$ in terms of flame thickness. The spatial resolution corresponds to approximately $15$ nodes per flame thickness. As studied in \citet{pizza_dynamics_2008-1} for the same channel width, the flame exhibits different dynamics depending on the inlet velocity. At low inlet velocity of about $u_{in}=10 \, \text{cm/s}$, periodic ignition and extinction of the flame is observed. 
The inlet velocity is then progressively increased until the oscillatory behaviour ceases and a stable flame can be sustained near the inlet of the channel. A further increase of the inlet velocity to $u_{in}=75 \, \text{cm/s}$ results in a symmetric ``V-shaped flame" in the channel, the flame being concave towards the unburnt mixture. Finally, at inflow higher than $u_{in}=165 \text{ cm/s}$, stable asymmetric flames are formed which shift downstream with increasing inlet velocity. The Reynolds number corresponding to the inlet velocity varies between $\text{Re}=5.36$ to $\text{Re}=88.52$, the reference length being the channel width and the reference viscosity corresponding to the viscosity at inlet composition and temperature. 
\subsubsection{Periodic ignition and extinction}
\begin{figure}
\includegraphics[width=1.0\linewidth]{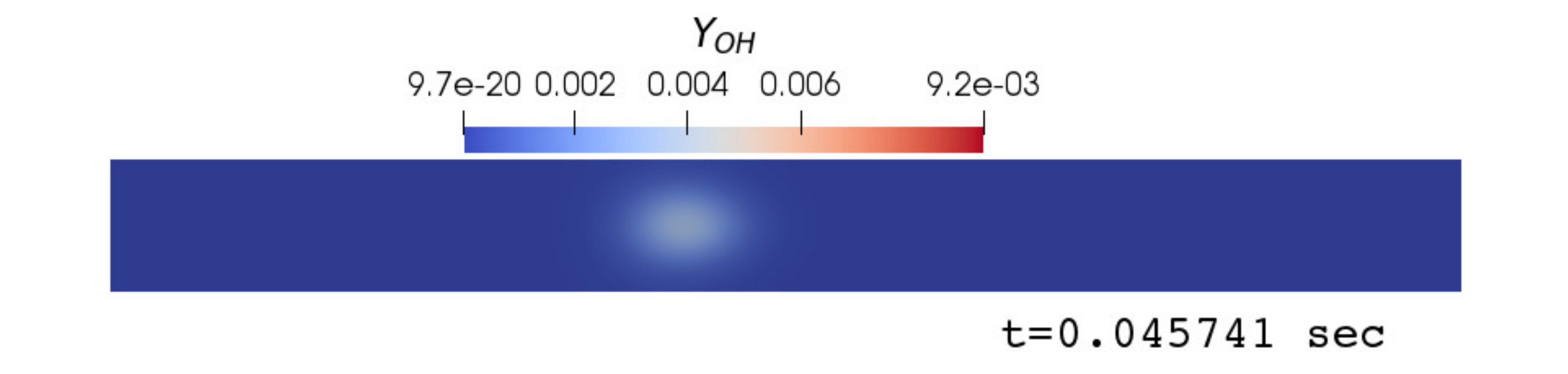} \\
\includegraphics[width=1.0\linewidth]{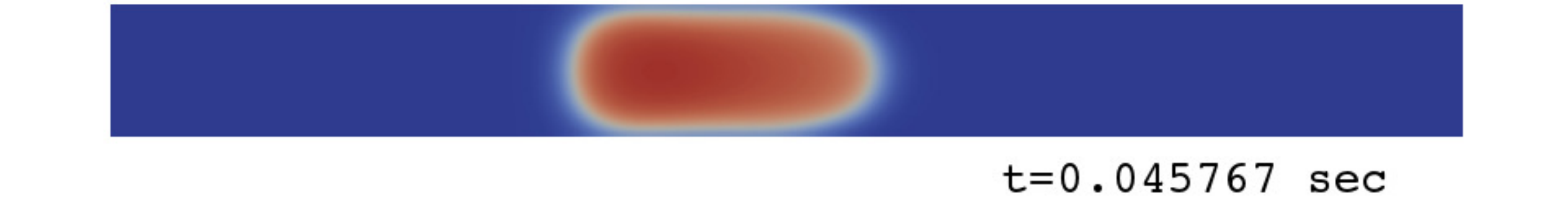}\\
\includegraphics[width=1.0\linewidth]{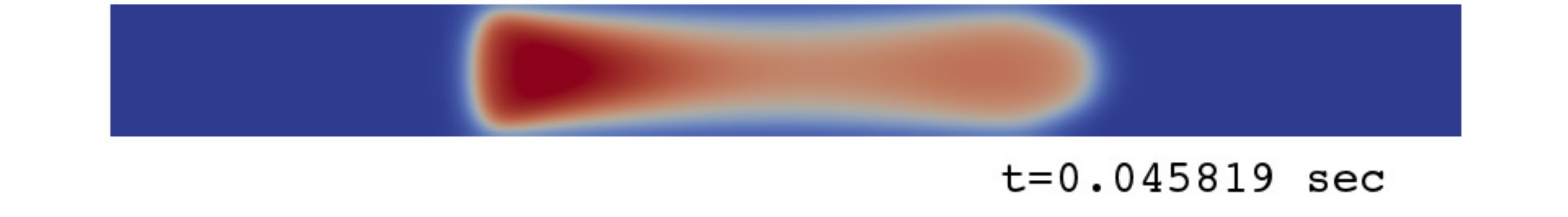}\\
\includegraphics[width=1.0\linewidth]{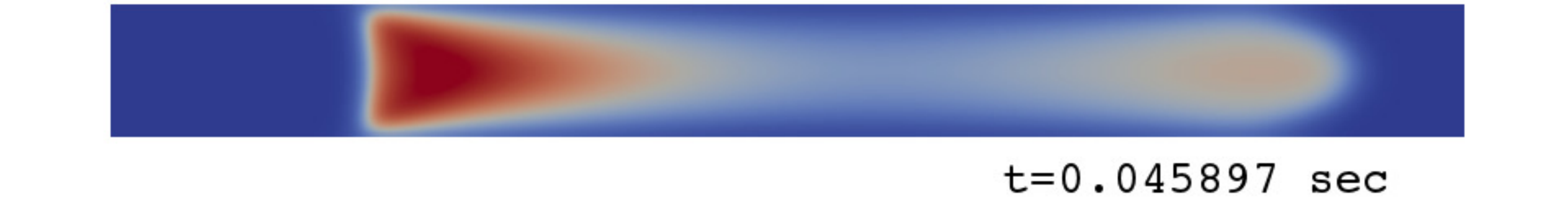}
\caption{Contours of the mass fraction of hydroxide \ce{OH} showing ignition, flame formation, splitting and propagation of the flame. Unburnt mixture enters the domain through the inlet on the left.}
\label{fig:igex}
\end{figure}

\begin{figure}
\includegraphics[width=1.0\linewidth]{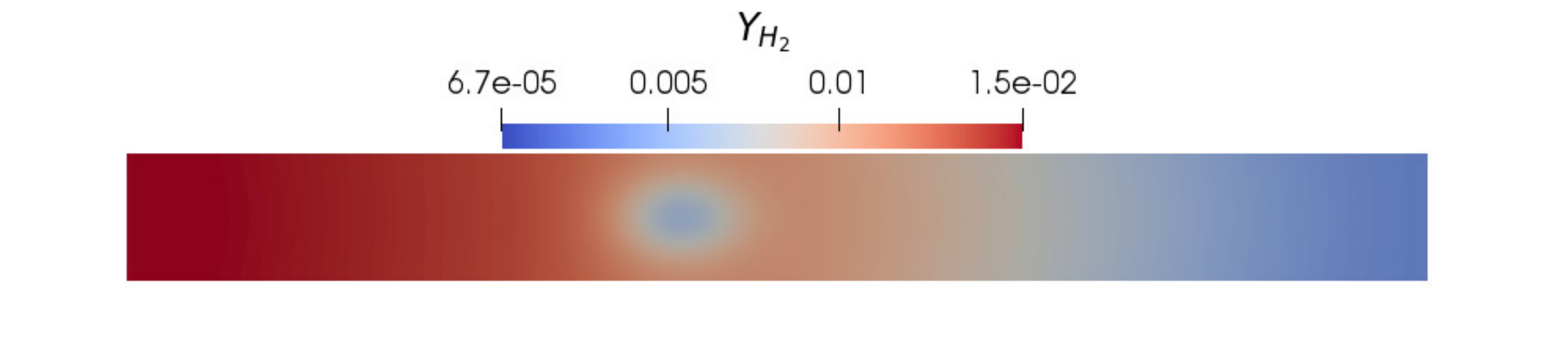}\\ 
\includegraphics[width=1.0\linewidth]{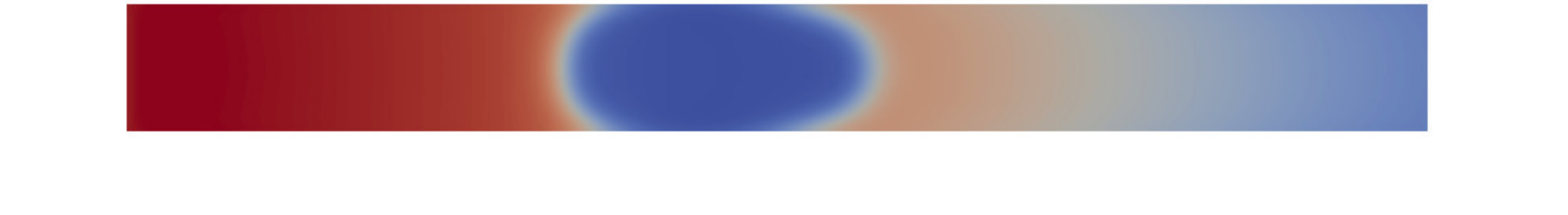}\\
\includegraphics[width=1.0\linewidth]{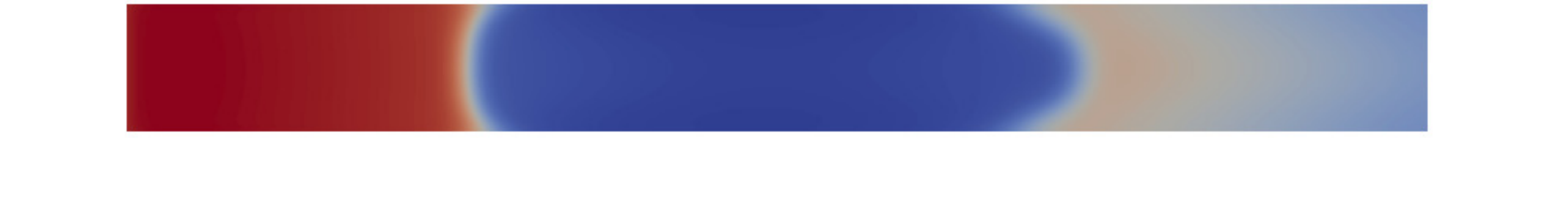}\\
\includegraphics[width=1.0\linewidth]{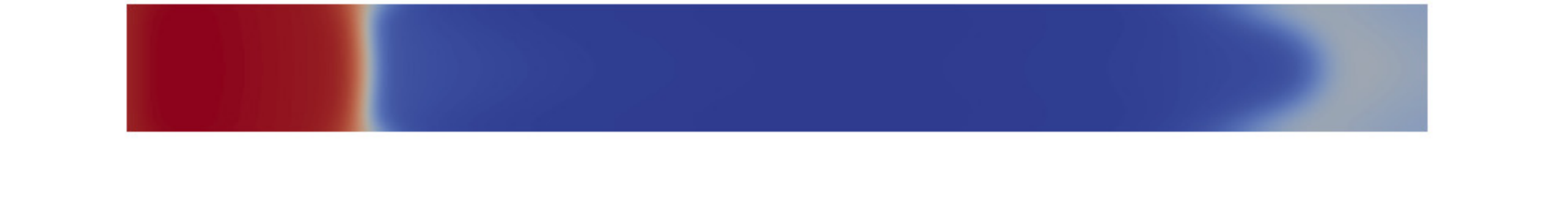}
\caption{Contours of the mass fraction of hydrogen \ce{H2} showing its consumption during the ignition-extinction process at time instants marked in the inset in Fig.\ \ref{fig:igex}. Unburnt mixture enters the domain through the inlet on the left.}
\label{fig:igex_H2}
\end{figure}

\begin{figure}
 \centering
 \includegraphics[keepaspectratio=true,width=0.95\textwidth]{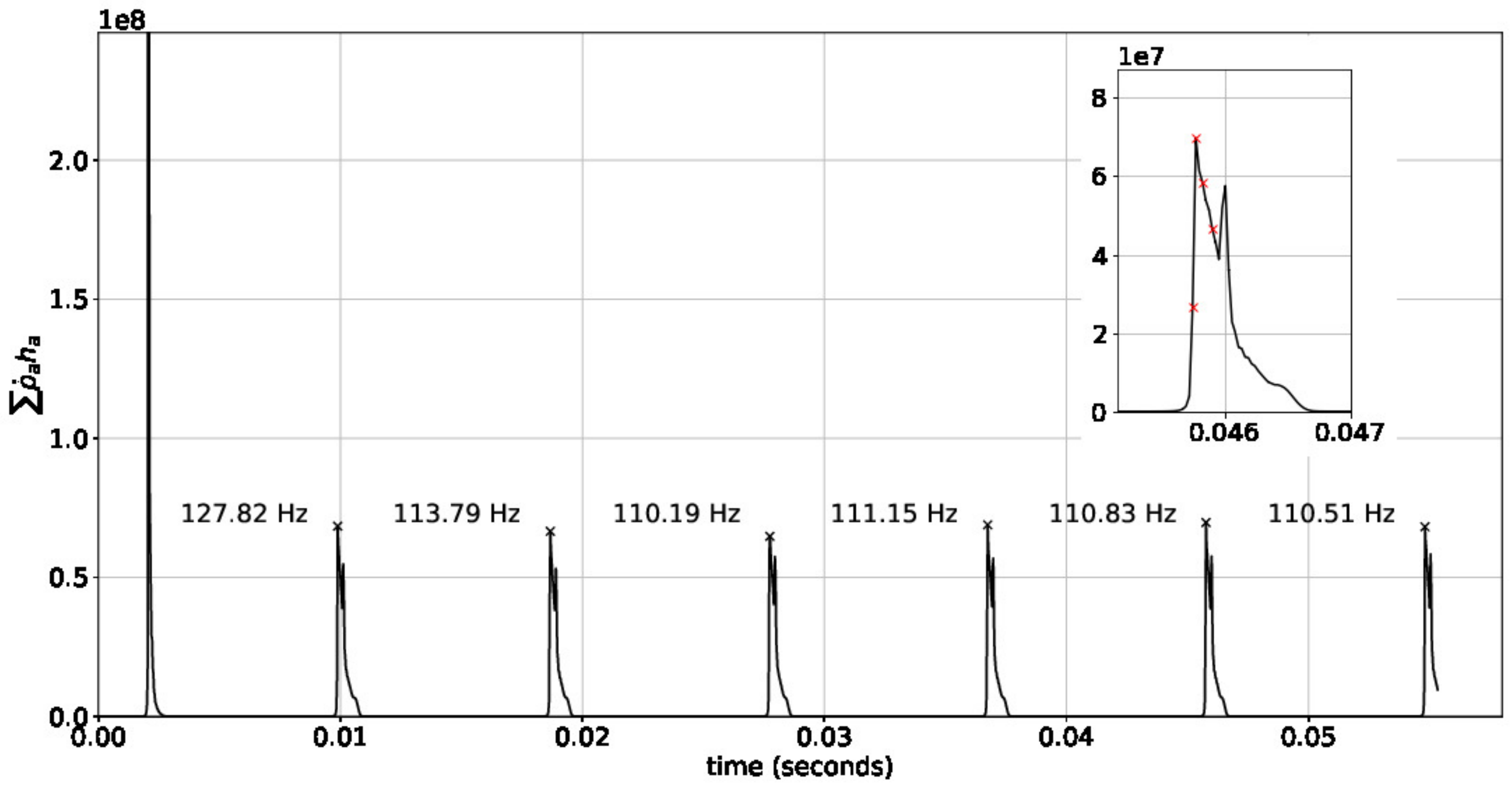}
  \caption{ Integrated heat release rate v/s time in the periodic ignition-extinction regime. The frequency from the LBM simulation is approximately $111\, \text{Hz}$, reference frequency from the DNS of \citet{pizza_dynamics_2008-1} is $106.9$ Hz. Time instants corresponding to Fig.\ \ref{fig:igex} are marked by red crosses in the inset figure.} 
 \label{fig:ihrr}
\end{figure}

The fluid in the bulk of the domain is initialized with the inlet unburnt composition and the inlet velocity is set to $u_{in}=10 \text{ cm/s}$. The initial temperature of the fluid in the bulk follows the wall temperature profile. As the fresh mixture passes between the heated walls, the reactants break into radicals which build up in the channel over time. This build up of radicals is associated with a long period of inactivity after which the mixture achieves a radical runaway and eventually a thermal runaway, leading to ignition. The mixture ignites at some distance downstream as seen in Fig.\ \ref{fig:igex}. The Fig.\ \ref{fig:igex} shows the hydroxide mass fraction which we will use as a marker to represent the ``flame" itself. The flame first forms a concentrated nearly circular structure which then propagates in both upstream as well as the downstream direction as visible in Fig. \ \ref{fig:igex}. This flame splitting occurs as the flame consumes the relatively fresh mixture in both possible directions. The frames in Figs.\ \ref{fig:igex_H2} show the mass fraction of hydrogen which is the deficient reactant. The flame then propagates and splits, consuming the deficient reactant in its path. The part of the flame travelling upstream is extinguished at the cold inlet whereas the part travelling downstream exits the channel through the outlet. Subsequently, the channel is again filled with the fresh mixture from the inlet and the process repeats periodically. In this regime, the maxima of all the species mass fractions as well as that of the temperature is located on the centreline of the channel. The behaviour is consistent with the DNS of \citet{pizza_dynamics_2008-1} and the subsequent simulations of \citet{alipoor_combustion_2016}. This phenomenon which is also referred to as a flame with repetitive extinction-ignition (FREI) has also been observed in methane/air combustion experiments of \citet{maruta_characteristics_2005} and numerical simulations of \citet{norton_combustion_2003}. The periodicity of the ignition-extinction behaviour has been presented through the variation of the integrated heat release rate with time in Fig.\ \ref{fig:ihrr}. In the figure, the heat release rate has been normalized with respect to the heat release rate of the unburnt state. The ignition events are seen produce a rise in the integrated heat release rate by $8$ orders of magnitude. The peaks are localized in time with an average frequency of approximately $111 \,\text{Hz}$. This is in good agreement with the frequency of $106.9 \,\text{Hz}$ reported in \citet{pizza_dynamics_2008-1}. Table \ref{tab:convergence} shows the convergence of the ignition-extinction frequency with resolution. The frequency changed by only $2.7 \%$ with an increase in the spatial resolution by $50 \%$. Therefore, the computations have been considered to be converged with respect to the resolution. The maximum velocity of the upstream propagation of the flame is found to be $16.8\, \text{m/s}$ and that of the downstream propagation is found to be $20\, \text{m/s}$ from the LBM simulations. For comparison, the maximum upstream propagation speed is reported to be $15\, \text{m/s}$ in \citet{pizza_dynamics_2008-1}. Overall, the LBM results quantitatively agree well with the DNS results. 

\begin{table}
\begin{center}
    \begin{tabular}{ l  l  l }
    \hline
    Domain size (nodes) & Resolution (nodes per $\delta_f$) & frequency (${\rm Hz}$) \\ \hline
    $150 \times 15$ & $6$ & $73$ \\ \hline
    $240 \times 24$ & $10$ & $108$ \\ \hline
    $360 \times 36$ & $15$ &  $111$ \\ \hline
    \end{tabular}
\end{center}
\caption{Frequency of the ignition-extinction phenomenon obtained from LBM simulations run with different spatial resolutions.}
\label{tab:convergence} 
\end{table}

\subsubsection{V-shaped stable flames}
\begin{figure}
	\centering
\includegraphics[width=1.0\linewidth]{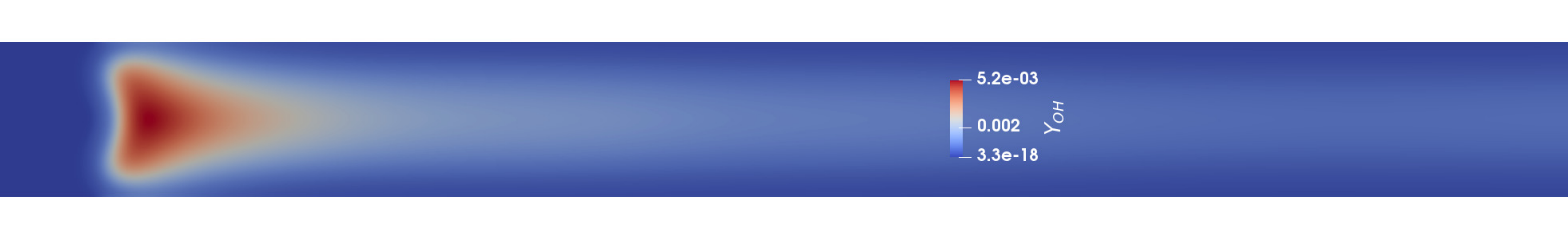}
\caption{V-shaped flames: Contours of hydroxide \ce{OH}  mass fraction. The flame is concave towards the unburnt mixture coming in from the left.}
\label{fig:vflame}
\end{figure}

\begin{figure}
	\centering
\includegraphics[width=1.0\linewidth]{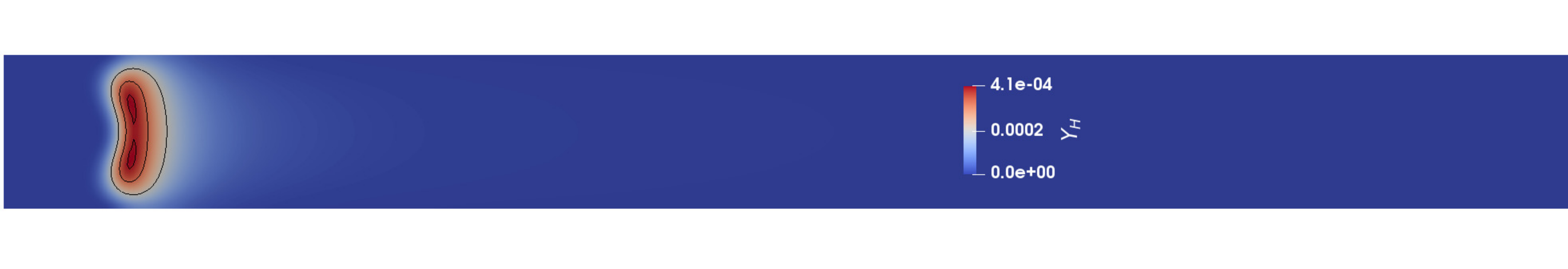} 
\caption{V-shaped flames: Contours of hydrogen radical \ce{H} mass fraction. Line contours highlight the shift of maxima away from the centreline.}
\label{fig:vflame_H}
\end{figure}

\begin{figure}
\centering
\includegraphics[width=1.0\linewidth]{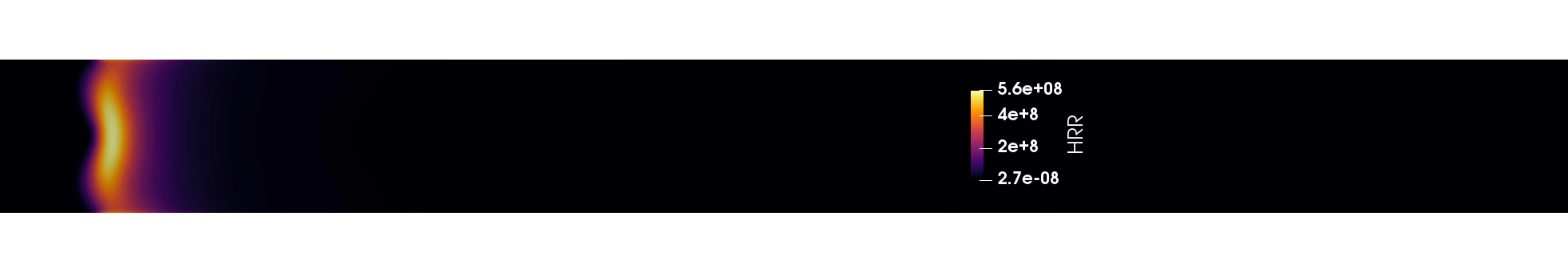} 
\caption{V-shaped flames: Contours of the heat release rate normalized by the heat release rate of the unburnt mixture.}
\label{fig:vflame_hrr}
\end{figure}

Using the solution from the ignition-extinction regime as an initial condition, the inlet velocity is progressively increased to $u_{in}=75\, \text{cm/s}$. In this regime, there is a sufficient flow of fresh mixture to sustain combustion and therefore a stable flame is formed in the channel. 
As evident in Fig.\ \ref{fig:vflame}, the flame assumes a ``V-shaped" structure which is concave towards the unburnt mixture. At this inlet velocity, the structure of all the species is symmetric about the centreline. The maxima of the mass fraction of all the species is located on the centreline except for the hydrogen radical. The hydrogen radial has a high molecular diffusivity, causing it to shift away from the channel centreline. This is evident from the line contours in Fig.\ \ref{fig:vflame_H}. The heat release rate contours in Fig.\ \ref{fig:vflame_hrr} show a localized heat release at the upstream interface of the flame. Also, the heat release rate contour follows a concave curvature that is similar to mass fraction contours of the hydroxide and the hydrogen radical. A maximum temperature of $1715\, \text{K}$ is attained in the flame. Shifting of the maxima of hydrogen at this inflow velocity as well as the concave shape of the flame is consistent with the findings of \citet{pizza_dynamics_2008-1}. With an increasing inlet velocity, the flame stabilizes further downstream from the inlet due to a relative increase in the difference between the velocity of the fresh mixture and the flame speed. Furthermore, at higher inlet velocities, more species shift away from the tube centreline. In a $3D$ microtube, the shifting of the maxima causes the flame to form a ring like structure \citep{pizza_three-dimensional_2010} around the tube centreline. We explore this phenomenon in detail in the corresponding $3D$ simulations. 

\subsubsection{Asymmetric stable flames}
\begin{figure}
 \centering
 \includegraphics[keepaspectratio=true,width=1.0\textwidth]{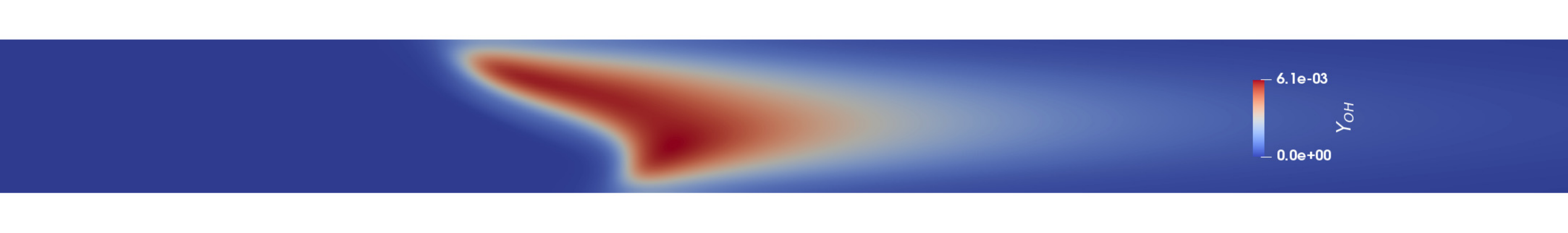}\\
 \includegraphics[keepaspectratio=true,width=1.0\textwidth]{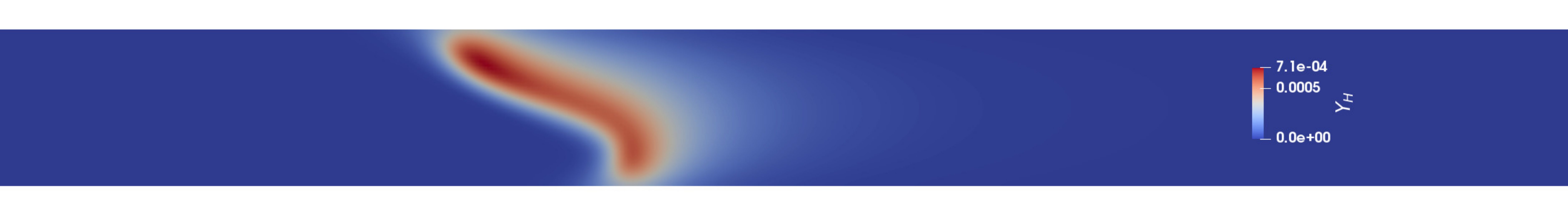}\\
 \includegraphics[keepaspectratio=true,width=1.0\textwidth]{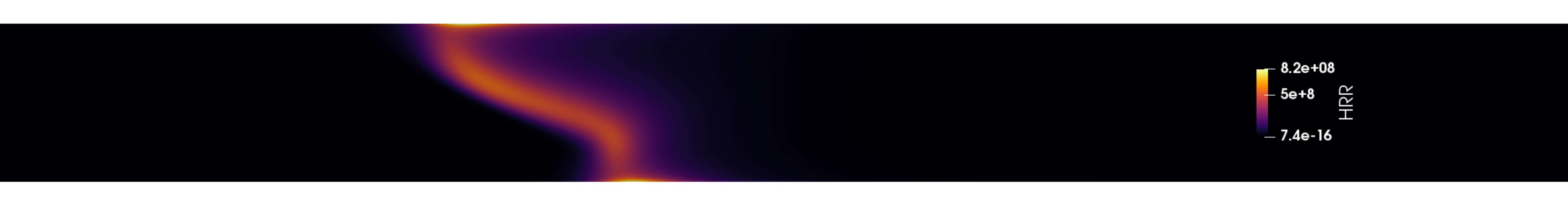}
  \caption{Lower asymmetric flame in the microchannel: Contours of hydroxide \ce{OH} mass fraction (top), hydrogen radical \ce{H} mass fraction (middle) and heat release rate (bottom).} 
 \label{fig:lower_asymmflame}
\end{figure}

\begin{figure}
 \centering
 \includegraphics[keepaspectratio=true,width=1.0\textwidth]{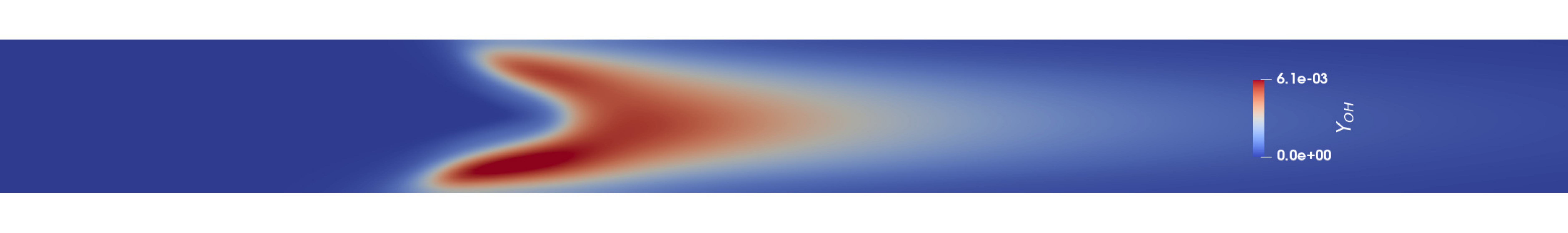}\\
 \includegraphics[keepaspectratio=true,width=1.0\textwidth]{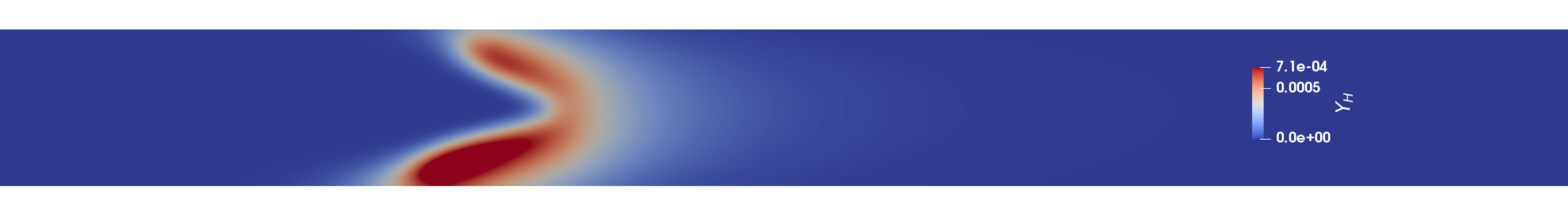}\\
 \includegraphics[keepaspectratio=true,width=1.0\textwidth]{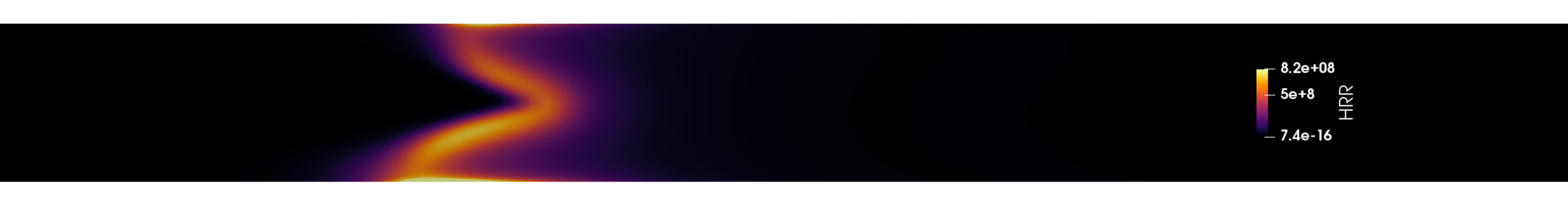}
  \caption{ Flame transitioning from a lower asymmetric shape to an upper asymmetric flame: Contours of hydroxide mass fraction (top), hydrogen mass fraction (middle) and heat release rate (bottom). } 
 \label{fig:trans_asymmflame}
\end{figure}

\begin{figure}
 \centering
 \includegraphics[keepaspectratio=true,width=1.0\textwidth]{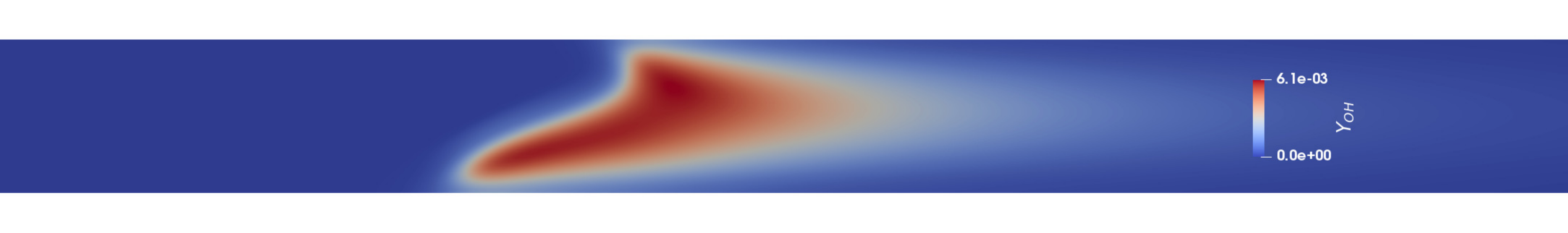}
 \includegraphics[keepaspectratio=true,width=1.0\textwidth]{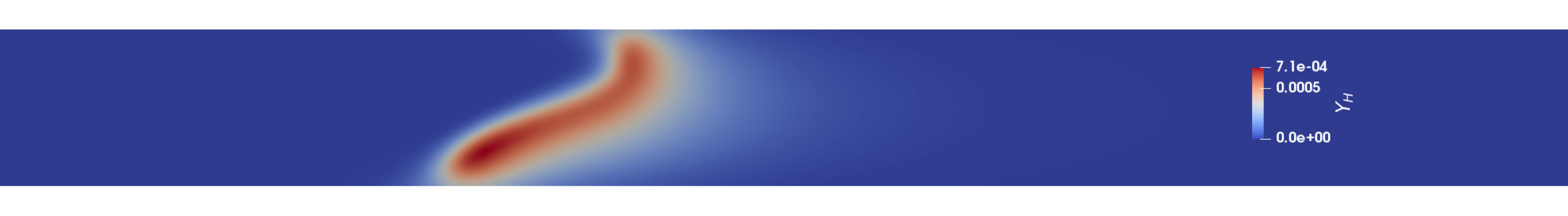}
 \includegraphics[keepaspectratio=true,width=1.0\textwidth]{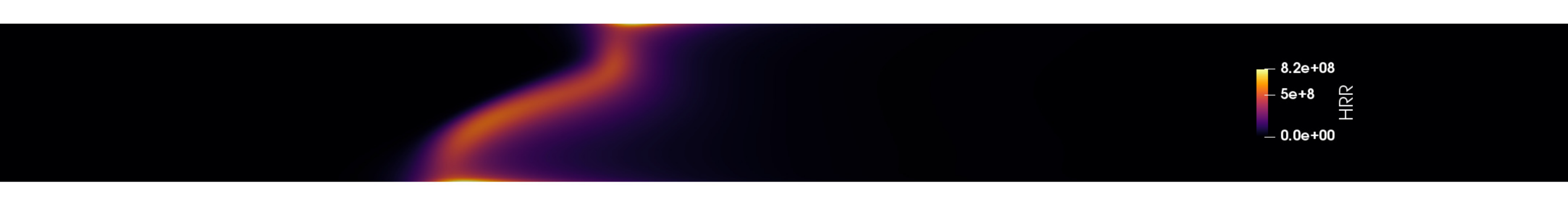}
  \caption{Upper asymmetric flame in the microchannel: Contours of hydroxide mass fraction (top), hydrogen mass fraction (middle) and heat release rate (bottom).} 
 \label{fig:upper_asymmflame}
\end{figure}

Starting from the V-shaped flame as an initial condition, we increase the inlet flow velocity gradually to $u_{in}=300\, \text{cm/s}$. After shifting downstream and maintaining its symmetric shape for some time, the flame transitions into an asymmetric stable flame. At the upstream interface between the flame and the unburnt mixture, a flame forming an acute angle with the lower wall is termed as a lower asymmetric flame  \citep{pizza_dynamics_2008-1}. Similarly, a flame forming an acute angle with the upper wall is termed an upper asymmetric flame. As shown in Fig. \ref{fig:lower_asymmflame}, a lower symmetric flame was first encountered in our computation. Interestingly, the asymmetric flame is metastable in this regime. The flame can be made to transition from a lower asymmetric shape to an upper asymmetric shape by heating the lower wall momentarily and then restoring the wall temperature back to the previous wall temperature of $960\, \text{K}$. A snapshot of the flame during this transition is shown in Fig.\ \ref{fig:trans_asymmflame}. An upper asymmetric flame formed as a result of the temperature perturbation is shown in Fig.\ \ref{fig:upper_asymmflame}. The resultant flame is also metastabe and remains in its upper asymmetric shape unless perturbed. The heat release rate profile is very similar to the profile of the mass fraction of hydrogen, except for the location of the maxima. The maxima of the heat release rate occurs at the walls in this regime. The distance of the flame from the inlet remains unchanged. In the LBM simulations, the location of the beginning of the flame is $0.24\,l$ from the inlet, which is in good agreement with $0.21\,l$, as obtained by the DNS of \citet{pizza_dynamics_2008-1}. The asymmetric nature of the flame and its metastable behaviour is consistent with the findings of \citet{pizza_dynamics_2008-1} for this regime. 
%
\subsection{Premixed hydrogen/air flame in microtube}
\label{sec:microtube}
For three-dimensional simulations, we choose the circular tube setup with a width $d=1.5\, \text{mm}$ from the simulations in \citet{pizza_three-dimensional_2010}. Richer dynamics are exhibited by this wide tube as compared to the narrower $1 \,\text{mm}$ tube. The composition of the incoming mixture is the same as for the two-dimensional simulations. The aspect ratio is also unchanged at $l/d=10$. The diameter corresponds to a flame thickness of $d=3.4884 \delta_f$. With the viscosity of the mixture at the inlet unburnt composition and at the inlet temperature as reference, the inlet velocity as the reference velocity and the diameter of the tube as the length scale, the Reynolds number is $\text{Re}=80.47$. We use a computational domain of $l \times d \times d=540 \times 54 \times 54$ nodes which translates into a spatial resolution of $15.5$ computational nodes per flame thickness. As discussed in section (\ref{sec:microchannel}) with the aid of table \ref{tab:convergence}, this resolution was sufficient to produce correct and converged results in the two-dimensional ignition-extinction simulations. In the three-dimensional simulations, we explore 
the `open axisymmetric flame' characterized by the maximum of the hydroxide radical being shifted away from the tube centreline. The iso-surfaces of hydroxide therefore form a ring shaped structure around the tube centreline. In the DNS of \citet{pizza_three-dimensional_2010}, such open flames are observed for an inflow velocity over two disconnected ranges, $60\, \text{cm/s} \leq u_{in} \leq 100\, \text{cm/s}$ and $170\, \text{cm/s} \leq u_{in} \leq 350\, \text{cm/s}$. In this work, we verify the existence of open flames by performing a simulation at an inflow of $u_{in}=100\, \text{cm/s}$.

\begin{figure}
 \centering
 \includegraphics[keepaspectratio=true,width=0.9\textwidth]{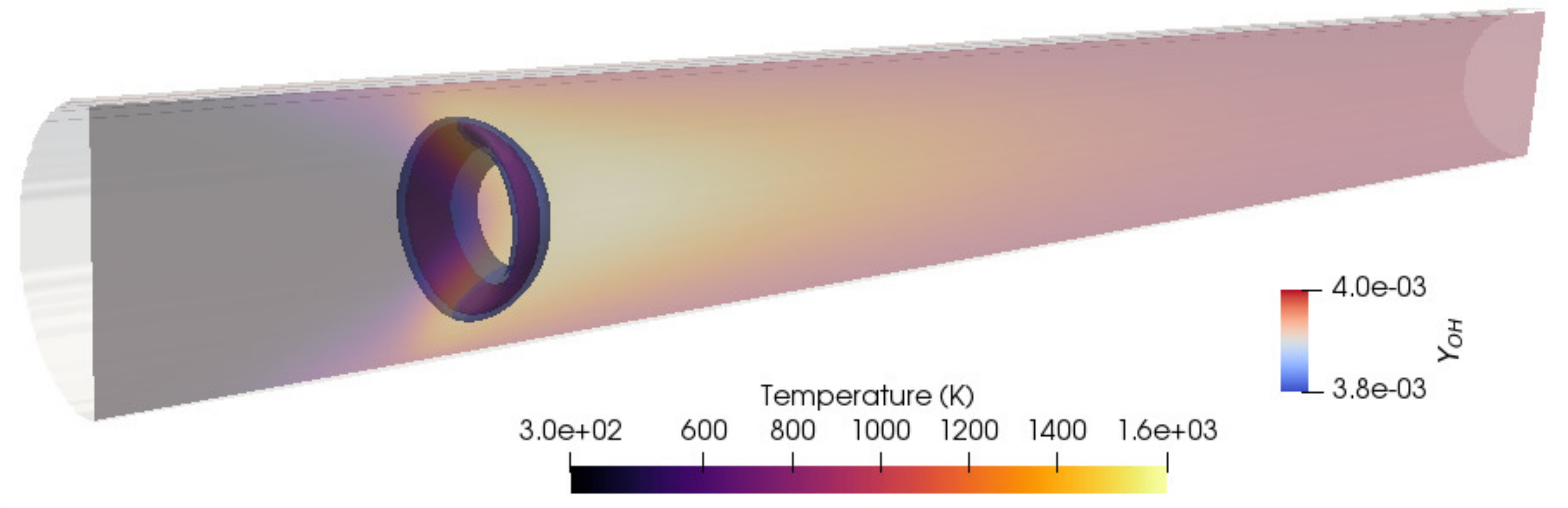}
  \caption{ Open flame in the microchannel: iso-surfaces of $Y_{OH}$ and slice of temperature contour at a $Z$ plane passing through the centre. }
 \label{fig:openflame}
\end{figure}

\begin{figure}
 \centering
 \includegraphics[keepaspectratio=true,width=0.9\textwidth]{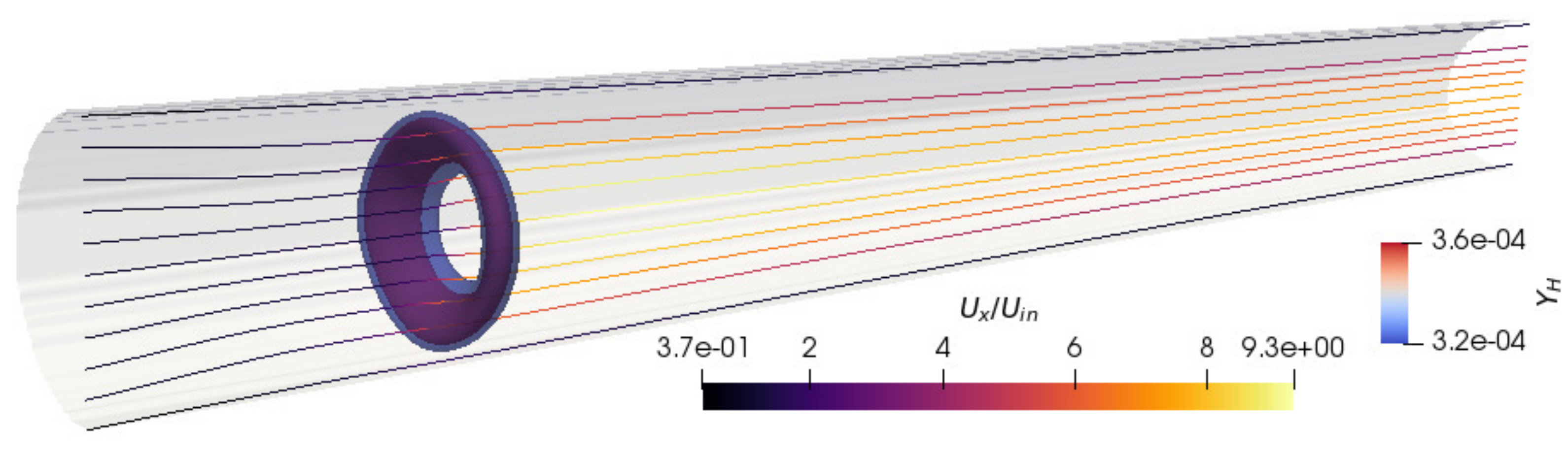}
  \caption{ Open flame in the microchannel: iso-surfaces of $Y_{H}$ and streamlines of the fluid velocity. } 
 \label{fig:openflame2}
\end{figure}

The bulk of the fluid in the tube is initialized with the inflow velocity $u_{in}=100\, \text{cm/s}$ and the temperature is initialized to the wall temperature profile. The composition is initialized with a 1D laminar flame solution computed for the same equivalence ratio as this 3D setup. The initial pressure in the domain is homogeneous at $1\, \text{atm}$. As a consequence of the initial conditions and the inflow velocity, the ignition-extinction regime is not encountered. The incoming fresh mixture enters the tube at a speed which is nearly twice of the flame speed. Therefore, the resulting flame does not propagate upstream into the inlet and stabilizes at a distance downstream of the inlet. The flame has the location of the maximum of the hydroxide radical and the temperature shifted away from the longitudinal axis of the tube. Therefore, as visible in Fig. \ref{fig:openflame}, iso-surfaces of the mass fraction of hydroxide form ring shaped structures. This type of a ring-like flame is  called an `open flame' in \citet{pizza_three-dimensional_2010}. At $u_{in}=100\, \text{cm/s}$, the open flame is axisymmetric and maintains a fixed distance of approximately $0.1 l$ from the inlet. The maximum temperature in the flame is $1649$ K. Fig. \ref{fig:openflame2} shows iso-surfaces of the mass fraction of the hydrogen radical, which also forms a ring structure similar to the hydroxide radical. Streamlines of the fluid velocity in Fig. \ref{fig:openflame2} show a marked acceleration in the fluid velocity downstream of the flame location. The post combustion fluid in the tube is seen to have attained a maximum velocity which is $9.3$ times of the inlet velocity. The maxima of the fluid velocity resides on the tube centreline.          

\section{Conclusion}
\label{sec:conclusion}
%

In this paper, we aimed at developing an accurate and robust LB model for reactive flows of practical interest.
In \citet{sawant_consistent_2021}, we proposed a lattice Boltzmann framework consisting of $M+2$ lattice Boltzmann equations for multicomponent mixtures of ideal gases. We introduced a new LBM system for the Stefan–Maxwell diffusion along with a reduced, mean-field description of the
mixture momentum and energy using a two-population approach. 
Thermodynamic consistency of this model allowed us to naturally account for the temperature and energy changes due to chemical reactions by including the energy of formation, which avoids any ad-hoc modelling for the heat of reaction. 
The proposed model uses the extended lattice Boltzmann method of \citet{saadat_extended_2021} for the mean fields and a multistep approach for integrating the mass source terms.
Furthermore, we introduced novel kinetic boundary conditions for walls, inlets and outlets that are compatible with the underlying reactive flow model.

Our model was validated in detail, starting from a zero-dimensional perfectly stirred reactor and the one-dimensional laminar flame.
The accuracy of the boundary conditions was assessed by simulations of premixed hydrogen/air flames in a microtube in both two and three dimensions. 
In all cases the results were found to be in excellent agreement with reference and DNS solutions that can be found in the literature.  

To conclude, the proposed model is not only a viable alternative to traditional reactive computational fluid dynamics but it is also the first model capable of solving reactive flows entirely in the lattice Boltzmann framework. We believe that this model marks a significant stage in the development of LBM by expanding the applicability of LBM to a wide range of setups including but not limited to diffusion, reactive flows and combustion. 

\noindent {\bf Acknowledgement}. 
This work was supported by European Research Council (ERC) Advanced Grant No. 834763-PonD. 
Computational resources at the Swiss National  Super  Computing  Center  CSCS  were  provided  under grant No. s1066.

\noindent {\bf Declaration of interests}.
The authors report no conflict of interest.


\bibliographystyle{jfm}
\bibliography{library}

\begin{thebibliography}{54}
\expandafter\ifx\csname natexlab\endcsname\relax\def\natexlab#1{#1}\fi
\def\au#1{#1} \def\ed#1{#1} \def\yr#1{#1}\def\at#1{#1}\def\jt#1{\textit{#1}}
  \def\bt#1{#1}\def\bvol#1{\textbf{#1}} \def\vol#1{#1} \def\pg#1{#1}
  \def\publ#1{#1}\def\arxiv#1{#1}\def\org#1{#1}\def\st#1{\textit{#1}}

\bibitem[Alipoor \& Mazaheri(2016)]{alipoor_combustion_2016}
{\sc \au{Alipoor, Alireza} \& \au{Mazaheri, Kiumars}} \yr{2016}  \at{Combustion
  characteristics and flame bifurcation in repetitive extinction-ignition
  dynamics for premixed hydrogen-air combustion in a heated micro channel}.
  \jt{Energy}  \bvol{109},  \pg{650--663}.

\bibitem[Bird {\em et~al.\/}(2007)Bird, Stewart \&
  Lightfoot]{bird_transport_2007}
{\sc \au{Bird, Robert~Byron}, \au{Stewart, Warren~E.} \& \au{Lightfoot,
  Edwin~N.}} \yr{2007} {\em Transport phenomena\/}, revised second edition edn.
   \publ{New York: John Wiley \& sons, inc}.

\bibitem[Boyd {\em et~al.\/}(2004)Boyd, Buick, Cosgrove \&
  Stansell]{boyd_application_2004}
{\sc \au{Boyd, J.}, \au{Buick, J.~M.}, \au{Cosgrove, J.~A.} \& \au{Stansell,
  P.}} \yr{2004}  \at{Application of the lattice {Boltzmann} method to arterial
  flow simulation: {Investigation} of boundary conditions for complex arterial
  geometries}.  \jt{Australasian Physics \& Engineering Sciences in Medicine}
  \bvol{27}~(4),  \pg{207--212}.

\bibitem[Chen \& Doolen(1998)]{chen_lattice_1998}
{\sc \au{Chen, S.} \& \au{Doolen, G.~D.}} \yr{1998}  \at{Lattice {Boltzmann}
  {Method} for {Fluid} {Flows}}.  \jt{Annu.{\textbackslash}
  Rev.{\textbackslash} Fluid Mech.}  \bvol{30},  \pg{329}.

\bibitem[Chikatamarla \& Karlin(2013)]{chikatamarla_entropic_2013}
{\sc \au{Chikatamarla, S.~S.} \& \au{Karlin, I.~V.}} \yr{2013}  \at{Entropic
  lattice {Boltzmann} method for turbulent flow simulations: {Boundary}
  conditions}.  \jt{Physica A}  \bvol{392},  \pg{1925--1930}.

\bibitem[Dorschner {\em et~al.\/}(2018)Dorschner, Bösch \&
  Karlin]{dorschner_particles_2018}
{\sc \au{Dorschner, B.}, \au{Bösch, F.} \& \au{Karlin, I.~V.}} \yr{2018}
  \at{Particles on demand for kinetic theory}.  \jt{Phys. Rev. Lett.}
  \bvol{121}~(13),  \pg{130602}.

\bibitem[Dorschner {\em et~al.\/}(2015)Dorschner, Chikatamarla, Bösch \&
  Karlin]{dorschner_grads_2015}
{\sc \au{Dorschner, B}, \au{Chikatamarla, Shyam~S}, \au{Bösch, Fabian} \&
  \au{Karlin, Iliya~V}} \yr{2015}  \at{Grad's approximation for moving and
  stationary walls in entropic lattice {Boltzmann} simulations}.  \jt{Journal
  of Computational Physics}  \bvol{295},  \pg{340--354}, publisher: Elsevier.

\bibitem[Dorschner {\em et~al.\/}(2016)Dorschner, Chikatamarla \&
  Karlin]{dorschner_entropic_2016}
{\sc \au{Dorschner, B.}, \au{Chikatamarla, S.~S.} \& \au{Karlin, I.~V.}}
  \yr{2016}  \at{Entropic {Lattice} {Boltzmann} {Method} for {Moving} and
  {Deforming} {Geometries} in {Three} {Dimensions}}.  \jt{Phys. Rev. E 95,
  063306 (2017)} ArXiv: http://arxiv.org/abs/1608.04658v1.

\bibitem[Dorschner {\em et~al.\/}(2017)Dorschner, Chikatamarla \&
  Karlin]{dorschner_transitional_2017}
{\sc \au{Dorschner, B.}, \au{Chikatamarla, S.~S.} \& \au{Karlin, I.~V.}}
  \yr{2017}  \at{Transitional flows with the entropic lattice {Boltzmann}
  method}.  \jt{J. Fluid Mech.}  \bvol{824},  \pg{388--412}.

\bibitem[Feng {\em et~al.\/}(2018)Feng, Tayyab \&
  Boivin]{feng_lattice-boltzmann_2018}
{\sc \au{Feng, Y.}, \au{Tayyab, M.} \& \au{Boivin, P.}} \yr{2018}  \at{A
  {Lattice}-{Boltzmann} model for low-{Mach} reactive flows}.  \jt{Combustion
  and Flame}  \bvol{196},  \pg{249--254}.

\bibitem[Frapolli {\em et~al.\/}(2016)Frapolli, Chikatamarla \&
  Karlin]{frapolli_entropic_2016}
{\sc \au{Frapolli, N.}, \au{Chikatamarla, S.~S.} \& \au{Karlin, I.~V.}}
  \yr{2016}  \at{Entropic lattice {Boltzmann} model for gas dynamics: {Theory},
  boundary conditions, and implementation}.  \jt{Physical Review E}
  \bvol{93}~(6),  \pg{063302}, publisher: APS.

\bibitem[Giovangigli(2012)]{giovangigli_multicomponent_2012}
{\sc \au{Giovangigli, V.}} \yr{2012} {\em Multicomponent flow modeling\/}.
  \publ{Birkhauser}.

\bibitem[Goodwin {\em et~al.\/}(2018)Goodwin, Speth, Moffat \&
  Weber]{goodwin_cantera_2018}
{\sc \au{Goodwin, D.~G.}, \au{Speth, R.~L.}, \au{Moffat, H.~K.} \& \au{Weber,
  B.~W.}} \yr{2018} {\em Cantera: {An} {Object}-oriented {Software} {Toolkit}
  for {Chemical} {Kinetics}, {Thermodynamics}, and {Transport} {Processes}\/}.

\bibitem[Guennebaud {\em et~al.\/}(2010)Guennebaud, Jacob \&
  {others}]{guennebaud_eigen_2010}
{\sc \au{Guennebaud, Gaël}, \au{Jacob, Benoît} \& \au{{others}}} \yr{2010}
  Eigen v3.

\bibitem[He {\em et~al.\/}(1998)He, Chen \& Doolen]{he_novel_1998}
{\sc \au{He, X.}, \au{Chen, S.} \& \au{Doolen, G.~D.}} \yr{1998}  \at{A novel
  thermal model for the lattice {Boltzmann} {Method} in incompressible limit}.
  \jt{J. Comp. Phys.}  \bvol{146}~(1),  \pg{282--300}.

\bibitem[Higuera \& Jiménez(1989)]{higuera_boltzmann_1989}
{\sc \au{Higuera, F.~J.} \& \au{Jiménez, J.}} \yr{1989}  \at{Boltzmann
  {Approach} to {Lattice} {Gas} {Simulations}}.  \jt{Europhysics Letters}
  \bvol{9}~(7),  \pg{663--668}, publisher: IOP Publishing.

\bibitem[Higuera \& Succi(1989)]{higuera_simulating_1989}
{\sc \au{Higuera, F.~J.} \& \au{Succi, S.}} \yr{1989}  \at{Simulating the
  {Flow} {Around} a {Circular} {Cylinder} with a {Lattice} {Boltzmann}
  {Equation}}.  \jt{EPL (Europhysics Letters)}  \bvol{8}~(6),  \pg{517}.

\bibitem[Hosseini {\em et~al.\/}(2020)Hosseini, Abdelsamie, Darabiha \&
  Thévenin]{hosseini_low-mach_2020}
{\sc \au{Hosseini, S.~A.}, \au{Abdelsamie, A.}, \au{Darabiha, N.} \&
  \au{Thévenin, D.}} \yr{2020}  \at{Low-{Mach} hybrid lattice
  {Boltzmann}-finite difference solver for combustion in complex flows}.
  \jt{Physics of Fluids}  \bvol{32}~(7),  \pg{077105}, \_eprint:
  https://doi.org/10.1063/5.0015034.

\bibitem[Hosseini {\em et~al.\/}(2018)Hosseini, Darabiha \&
  Thévenin]{hosseini_mass-conserving_2018}
{\sc \au{Hosseini, S.~A.}, \au{Darabiha, N.} \& \au{Thévenin, D.}} \yr{2018}
  \at{Mass-conserving advection–diffusion lattice {Boltzmann} model for
  multi-species reacting flows}.  \jt{Physica A: Statistical Mechanics and its
  Applications}  \bvol{499},  \pg{40 -- 57}.

\bibitem[Hosseini {\em et~al.\/}(2019)Hosseini, Safari, Darabiha, Thévenin \&
  Krafczyk]{hosseini_hybrid_2019}
{\sc \au{Hosseini, S.~A.}, \au{Safari, H.}, \au{Darabiha, N.}, \au{Thévenin,
  D.} \& \au{Krafczyk, M.}} \yr{2019}  \at{Hybrid lattice {Boltzmann} - finite
  difference model for low {Mach} number combustion simulation}.
  \jt{Combustion and Flame}  \bvol{209},  \pg{394--404}.

\bibitem[Karlin {\em et~al.\/}(2013)Karlin, Sichau \&
  Chikatamarla]{karlin_consistent_2013}
{\sc \au{Karlin, I.~V.}, \au{Sichau, D.} \& \au{Chikatamarla, S.~S.}} \yr{2013}
   \at{Consistent two-population lattice {Boltzmann} model for thermal flows}.
  \jt{Phys. Rev. E}  \bvol{88}~(6),  \pg{063310}, publisher: American Physical
  Society.

\bibitem[Kee {\em et~al.\/}(2003)Kee, Coltrin \& Glarborg]{kee_chemically_2003}
{\sc \au{Kee, R.~J.}, \au{Coltrin, M.~E.} \& \au{Glarborg, P.}} \yr{2003} {\em
  Chemically {Reacting} {Flow}: {Theory} and {Practice}\/}.  \publ{Hoboken, NJ:
  Wiley Pub. Co.}

\bibitem[Ladd(1994)]{ladd_numerical_1994}
{\sc \au{Ladd, Anthony J.~C.}} \yr{1994}  \at{Numerical simulations of
  particulate suspensions via a discretized {Boltzmann} equation. {Part} 1.
  {Theoretical} foundation}.  \jt{Journal of Fluid Mechanics}  \bvol{271},
  \pg{285--309}.

\bibitem[Li {\em et~al.\/}(2004)Li, Zhao, Kazakov \& Dryer]{li_updated_2004}
{\sc \au{Li, Juan}, \au{Zhao, Zhenwei}, \au{Kazakov, Andrei} \& \au{Dryer,
  Frederick~L.}} \yr{2004}  \at{An updated comprehensive kinetic model of
  hydrogen combustion}.  \jt{International Journal of Chemical Kinetics}
  \bvol{36}~(10),  \pg{566--575}.

\bibitem[Lin \& Luo(2018)]{lin_mrt_2018}
{\sc \au{Lin, C.} \& \au{Luo, K.~H.}} \yr{2018}  \at{{MRT} discrete {Boltzmann}
  method for compressible exothermic reactive flows}.  \jt{Computers \& Fluids}
   \bvol{166},  \pg{176--183}.

\bibitem[Maruta {\em et~al.\/}(2005)Maruta, Kataoka, Kim, Minaev \&
  Fursenko]{maruta_characteristics_2005}
{\sc \au{Maruta, Kaoru}, \au{Kataoka, Takuya}, \au{Kim, Nam~Il}, \au{Minaev,
  Sergey} \& \au{Fursenko, Roman}} \yr{2005}  \at{Characteristics of combustion
  in a narrow channel with a temperature gradient}.  \jt{Proceedings of the
  Combustion Institute}  \bvol{30}~(2),  \pg{2429--2436}.

\bibitem[Mathur {\em et~al.\/}(1967)Mathur, Tondon \&
  Saxena]{mathur_thermal_1967}
{\sc \au{Mathur, S.}, \au{Tondon, P.~K.} \& \au{Saxena, S.~C.}} \yr{1967}
  \at{Thermal conductivity of binary, ternary and quaternary mixtures of rare
  gases}.  \jt{Molecular Physics}  \bvol{12}~(6),  \pg{569--579}, publisher:
  Taylor and Francis.

\bibitem[Mazloomi {\em et~al.\/}(2015)Mazloomi, Chikatamarla \&
  Karlin]{mazloomi_entropic_2015}
{\sc \au{Mazloomi, A.~M.}, \au{Chikatamarla, S.~S.} \& \au{Karlin, I.~V.}}
  \yr{2015}  \at{Entropic lattice {Boltzmann} method for multiphase flows}.
  \jt{Phys. Rev. Lett.}  \bvol{114}~(17),  \pg{174502}.

\bibitem[Mazloomi {\em et~al.\/}(2017)Mazloomi, Chikatamarla \&
  Karlin]{mazloomi_drops_2017}
{\sc \au{Mazloomi, A.~M.}, \au{Chikatamarla, S.~S.} \& \au{Karlin, I.~V.}}
  \yr{2017}  \at{Drops bouncing off macro-textured superhydrophobic surfaces}.
  \jt{J. Fluid Mech.}  \bvol{824},  \pg{866--885}.

\bibitem[Montemore {\em et~al.\/}(2017)Montemore, Montessori, Succi, Barroo,
  Falcucci, Bell \& Kaxiras]{montemore_effect_2017}
{\sc \au{Montemore, Matthew~M.}, \au{Montessori, Andrea}, \au{Succi, Sauro},
  \au{Barroo, Cédric}, \au{Falcucci, Giacomo}, \au{Bell, David~C.} \&
  \au{Kaxiras, Efthimios}} \yr{2017}  \at{Effect of nanoscale flows on the
  surface structure of nanoporous catalysts}.  \jt{The Journal of Chemical
  Physics}  \bvol{146}~(21),  \pg{214703}.

\bibitem[Montessori {\em et~al.\/}(2016)Montessori, Prestininzi, Rocca,
  Falcucci, Succi \& Kaxiras]{montessori_effects_2016}
{\sc \au{Montessori, A.}, \au{Prestininzi, P.}, \au{Rocca, M.~La},
  \au{Falcucci, G.}, \au{Succi, S.} \& \au{Kaxiras, E.}} \yr{2016}  \at{Effects
  of {Knudsen} diffusivity on the effective reactivity of nanoporous catalyst
  media}.  \jt{Journal of Computational Science}  \bvol{17},  \pg{377 -- 383}.

\bibitem[Mott-Smith(1951)]{mott-smith_solution_1951}
{\sc \au{Mott-Smith, H.~M.}} \yr{1951}  \at{The {Solution} of the {Boltzmann}
  {Equation} for a {Shock} {Wave}}.  \jt{Physical Review}  \bvol{82}~(6),
  \pg{885--892}.

\bibitem[Norton \& Vlachos(2003)]{norton_combustion_2003}
{\sc \au{Norton, D.G.} \& \au{Vlachos, D.G.}} \yr{2003}  \at{Combustion
  characteristics and flame stability at the microscale: a {CFD} study of
  premixed methane/air mixtures}.  \jt{Chemical Engineering Science}
  \bvol{58}~(21),  \pg{4871--4882}.

\bibitem[Pizza {\em et~al.\/}(2008{\natexlab{{\em a\/}}})Pizza, Frouzakis,
  Mantzaras, Tomboulides \& Boulouchos]{pizza_dynamics_2008}
{\sc \au{Pizza, Gianmarco}, \au{Frouzakis, Christos~E.}, \au{Mantzaras, John},
  \au{Tomboulides, Ananias~G.} \& \au{Boulouchos, Konstantinos}}
  \yr{2008{\natexlab{{\em a\/}}}}  \at{Dynamics of premixed hydrogen/air flames
  in mesoscale channels}.  \jt{Combustion and Flame}  \bvol{155}~(1-2),
  \pg{2--20}.

\bibitem[Pizza {\em et~al.\/}(2008{\natexlab{{\em b\/}}})Pizza, Frouzakis,
  Mantzaras, Tomboulides \& Boulouchos]{pizza_dynamics_2008-1}
{\sc \au{Pizza, Gianmarco}, \au{Frouzakis, Christos~E.}, \au{Mantzaras, John},
  \au{Tomboulides, Ananias~G.} \& \au{Boulouchos, Konstantinos}}
  \yr{2008{\natexlab{{\em b\/}}}}  \at{Dynamics of premixed hydrogen/air flames
  in microchannels}.  \jt{Combustion and Flame}  \bvol{152}~(3),
  \pg{433--450}.

\bibitem[Pizza {\em et~al.\/}(2010)Pizza, Frouzakis, Mantzaras, Tomboulides \&
  Boulouchos]{pizza_three-dimensional_2010}
{\sc \au{Pizza, G.}, \au{Frouzakis, C.~E.}, \au{Mantzaras, J.},
  \au{Tomboulides, A.~G.} \& \au{Boulouchos, K.}} \yr{2010}
  \at{Three-dimensional simulations of premixed hydrogen/air flames in
  microtubes}.  \jt{Journal of Fluid Mechanics}  \bvol{658},  \pg{463--491},
  publisher: Cambridge University Press.

\bibitem[Poinsot \& Lele(1992)]{poinsot_boundary_1992}
{\sc \au{Poinsot, T.J} \& \au{Lele, S.K}} \yr{1992}  \at{Boundary conditions
  for direct simulations of compressible viscous flows}.  \jt{Journal of
  Computational Physics}  \bvol{101}~(1),  \pg{104--129}.

\bibitem[Poinsot \& Veynante(2005)]{poinsot_theoretical_2005}
{\sc \au{Poinsot, T.} \& \au{Veynante, D.}} \yr{2005} {\em Theoretical and
  numerical combustion\/}.  \publ{R.T. Edwards, Inc.}

\bibitem[Saadat {\em et~al.\/}(2019)Saadat, Bösch \&
  Karlin]{saadat_lattice_2019}
{\sc \au{Saadat, M.~H.}, \au{Bösch, F.} \& \au{Karlin, I.~V.}} \yr{2019}
  \at{Lattice {Boltzmann} model for compressible flows on standard lattices:
  {Variable} {Prandtl} number and adiabatic exponent}.  \jt{Phys. Rev. E}
  \bvol{99}~(1),  \pg{013306}, publisher: American Physical Society.

\bibitem[Saadat {\em et~al.\/}(2021)Saadat, Dorschner \&
  Karlin]{saadat_extended_2021}
{\sc \au{Saadat, Mohammad~Hossein}, \au{Dorschner, Benedikt} \& \au{Karlin,
  Ilya}} \yr{2021}  \at{Extended {Lattice} {Boltzmann} {Model}}.  \jt{Entropy}
  \bvol{23}~(4).

\bibitem[Sawant {\em et~al.\/}(2021{\natexlab{{\em a\/}}})Sawant, Dorschner \&
  Karlin]{sawant_consistent_2021}
{\sc \au{Sawant, N.}, \au{Dorschner, B.} \& \au{Karlin, I.~V.}}
  \yr{2021{\natexlab{{\em a\/}}}}  \at{Consistent lattice {Boltzmann} model for
  multicomponent mixtures}.  \jt{Journal of Fluid Mechanics}  \bvol{909},
  \pg{A1}.

\bibitem[Sawant {\em et~al.\/}(2021{\natexlab{{\em b\/}}})Sawant, Dorschner \&
  Karlin]{sawant_lattice_2021}
{\sc \au{Sawant, N}, \au{Dorschner, B} \& \au{Karlin, I~V}}
  \yr{2021{\natexlab{{\em b\/}}}}  \at{A lattice {Boltzmann} model for reactive
  mixtures}.  \jt{Phil. Trans. R. Soc. A}  \bvol{379},  \pg{15}.

\bibitem[Shan {\em et~al.\/}(2006)Shan, Yuan \& Chen]{shan_kinetic_2006}
{\sc \au{Shan, X.}, \au{Yuan, X.~F.} \& \au{Chen, H.}} \yr{2006}  \at{Kinetic
  theory representation of hydrodynamics: a way beyond the {Navier}–{Stokes}
  equation}.  \jt{J. Fluid Mech.}  \bvol{550},  \pg{413--441}.

\bibitem[Smith {\em et~al.\/}(1999)Smith, Golden, Frenklach, Moriarty,
  Eiteneer, Goldenberg, Bowman, Hanson, Song, Gardiner, Lissianski \&
  Qin]{smith_gri-mech_1999}
{\sc \au{Smith, G.~P.}, \au{Golden, D.~M.}, \au{Frenklach, M.}, \au{Moriarty,
  N.~W.}, \au{Eiteneer, B.}, \au{Goldenberg, M.}, \au{Bowman, C.~T.},
  \au{Hanson, R.~K.}, \au{Song, S.}, \au{Gardiner, Jr. W.~C.}, \au{Lissianski,
  V.~V.} \& \au{Qin, Z.}} \yr{1999} {\em {GRI}-{Mech} 3.0\/}.

\bibitem[Tayyab {\em et~al.\/}(2020)Tayyab, Radisson, Almarcha, Denet \&
  Boivin]{tayyab_experimental_2020}
{\sc \au{Tayyab, Muhammad}, \au{Radisson, Basile}, \au{Almarcha, Christophe},
  \au{Denet, Bruno} \& \au{Boivin, Pierre}} \yr{2020}  \at{Experimental and
  numerical {Lattice}-{Boltzmann} investigation of the {Darrieus}–{Landau}
  instability}.  \jt{Combustion and Flame}  \bvol{221},  \pg{103--109}.

\bibitem[Tayyab {\em et~al.\/}(2021)Tayyab, Zhao \&
  Boivin]{tayyab_lattice-boltzmann_2021}
{\sc \au{Tayyab, M.}, \au{Zhao, S.} \& \au{Boivin, P.}} \yr{2021}
  \at{Lattice-{Boltzmann} modeling of a turbulent bluff-body stabilized flame}.
   \jt{Physics of Fluids}  \bvol{33}~(3),  \pg{031701}.

\bibitem[Thampi {\em et~al.\/}(2013)Thampi, Ansumali, Adhikari \&
  Succi]{thampi_isotropic_2013}
{\sc \au{Thampi, S.~P.}, \au{Ansumali, S.}, \au{Adhikari, R.} \& \au{Succi,
  S.}} \yr{2013}  \at{Isotropic discrete {Laplacian} operators from lattice
  hydrodynamics}.  \jt{Journal of Computational Physics}  \bvol{234},  \pg{1 --
  7}.

\bibitem[Wilke(1950)]{wilke_viscosity_1950}
{\sc \au{Wilke, C.~R.}} \yr{1950}  \at{A {Viscosity} {Equation} for {Gas}
  {Mixtures}}.  \jt{The Journal of Chemical Physics}  \bvol{18}~(4),
  \pg{517--519}.

\bibitem[Williams(1985)]{williams_combustion_1985}
{\sc \au{Williams, F.~A.}} \yr{1985} {\em Combustion theory: the fundamental
  theory of chemically reacting flow systems\/}.  \publ{Redwood City, Calif.:
  Benjamin/Cummings Pub. Co.}

\bibitem[Wöhrwag {\em et~al.\/}(2018)Wöhrwag, Semprebon, Mazloomi, Karlin \&
  Kusumaatmaja]{wohrwag_ternary_2018}
{\sc \au{Wöhrwag, M.}, \au{Semprebon, C.}, \au{Mazloomi, A.~M.}, \au{Karlin,
  I.} \& \au{Kusumaatmaja, H.}} \yr{2018}  \at{Ternary free-energy entropic
  lattice {Boltzmann} model with a high density ratio}.  \jt{Phys. Rev. Lett.}
  \bvol{120}~(23),  \pg{234501}.

\bibitem[Xu \& Sagaut(2013)]{xu_analysis_2013}
{\sc \au{Xu, Hui} \& \au{Sagaut, Pierre}} \yr{2013}  \at{Analysis of the
  absorbing layers for the weakly-compressible lattice {Boltzmann} methods}.
  \jt{Journal of Computational Physics}  \bvol{245},  \pg{14--42}, publisher:
  Elsevier.

\bibitem[Yan {\em et~al.\/}(2013)Yan, Xu, Zhang, Ying \& Li]{yan_lattice_2013}
{\sc \au{Yan, B.}, \au{Xu, A.~G.}, \au{Zhang, G.~C.}, \au{Ying, Y.~J.} \&
  \au{Li, H.}} \yr{2013}  \at{Lattice {Boltzmann} model for combustion and
  detonation}.  \jt{Frontiers of Physics}  \bvol{8}~(1),  \pg{94--110}.

\bibitem[Yang {\em et~al.\/}(2018)Yang, West \& Harris]{yang_coupled_2018}
{\sc \au{Yang, H.~Q.}, \au{West, Jeff} \& \au{Harris, Robert~E.}} \yr{2018}
  \at{Coupled {Fluid}–{Structure} {Interaction} {Analysis} of {Solid}
  {Rocket} {Motor} with {Flexible} {Inhibitors}}.  \jt{Journal of Spacecraft
  and Rockets}  \bvol{55}~(2),  \pg{303--314}, \_eprint:
  https://doi.org/10.2514/1.A33947.

\bibitem[Ziegler(1993)]{ziegler_boundary_1993}
{\sc \au{Ziegler, Donald~P.}} \yr{1993}  \at{Boundary conditions for lattice
  {Boltzmann} simulations}.  \jt{Journal of Statistical Physics}
  \bvol{71}~(5-6),  \pg{1171--1177}.

\end{thebibliography}

\end{document}